\documentclass[twocolumn,twocolappendix]{aastex702}

\makeatletter
\def\email@error#1{}
\makeatother

\graphicspath{{./}}
\shorttitle{Cool Circumgalactic Gas across the SHMR}
\shortauthors{Chen et al.}
\journalinfo{}

\begin{document}

\title{The Cool Circumgalactic Echo of the Stellar-to-Halo Mass Relation}

\author[0009-0004-5989-6005]{Zeyu Chen}
\affiliation{Department of Astronomy, University of Science and Technology of China, Hefei 230026, China}
\affiliation{School of Astronomy and Space Science, University of Science and Technology of China, Hefei 230026, China}

\author[0000-0003-1588-9394]{Enci Wang}\thanks{Email: \href{mailto:ecwang16@ustc.edu.cn}{ecwang16@ustc.edu.cn}}
\affiliation{Department of Astronomy, University of Science and Technology of China, Hefei 230026, China}
\affiliation{School of Astronomy and Space Science, University of Science and Technology of China, Hefei 230026, China}

\author[0000-0002-4911-6990]{Huiyuan Wang}\thanks{Email: \href{mailto:whywang@ustc.edu.cn}{whywang@ustc.edu.cn}}
\affiliation{Department of Astronomy, University of Science and Technology of China, Hefei 230026, China}
\affiliation{School of Astronomy and Space Science, University of Science and Technology of China, Hefei 230026, China}

\author[0000-0001-5356-2419]{Houjun Mo}
\affiliation{Department of Astronomy, University of Massachusetts, Amherst, MA 01003-9305, USA}

\author[0000-0001-6947-5846]{Luis C. Ho}
\affiliation{Kavli Institute for Astronomy and Astrophysics, Peking University, Beijing, China}
\affiliation{Department of Astronomy, School of Physics, Peking University, Beijing, China}

\author[0009-0008-1319-498X]{Haoran Yu}
\affiliation{Department of Astronomy, University of Science and Technology of China, Hefei 230026, China}
\affiliation{School of Astronomy and Space Science, University of Science and Technology of China, Hefei 230026, China}

\author[0000-0002-2941-646X]{Zhijie Qu}
\affiliation{Department of Astronomy, Tsinghua University, Beijing, China}

\author[0000-0002-6684-3997]{Hu Zou}
\affiliation{National Astronomical Observatories, Chinese Academy of Sciences, Beijing 100012, China}

\author[0000-0002-3983-6484]{Siwei Zou}
\affiliation{Chinese Academy of Sciences South America Center for Astronomy, National Astronomical Observatories, CAS, Beijing 100101, China}
\affiliation{Departamento de Astronom\'ia, Universidad de Chile, Santiago, Chile}

\author[0000-0003-3667-1060]{Zhicheng He}
\affiliation{Department of Astronomy, University of Science and Technology of China, Hefei 230026, China}
\affiliation{School of Astronomy and Space Science, University of Science and Technology of China, Hefei 230026, China}

\author[0009-0000-7307-6362]{Cheqiu Lyu}
\affiliation{Department of Astronomy, University of Science and Technology of China, Hefei 230026, China}
\affiliation{School of Astronomy and Space Science, University of Science and Technology of China, Hefei 230026, China}

\author[0009-0006-7343-8013]{Chengyu Ma}
\affiliation{Department of Astronomy, University of Science and Technology of China, Hefei 230026, China}
\affiliation{School of Astronomy and Space Science, University of Science and Technology of China, Hefei 230026, China}

\author[0009-0000-2569-2742]{Weiwen Kong}
\affiliation{School of Mechanics and Photoelectric Physics, Anhui University of Science and Technology, Huainan 232001, China}

\begin{abstract}
The stellar-to-halo mass relation (SHMR), a key prediction of galaxy formation models, peaks at $M_* \sim 3\times10^{10}\,\mathrm{M_\odot}$. This scale is often attributed to circumgalactic baryonic feedback processes, however, direct observational evidence linking the two remains scarce. Using Dark Energy Spectroscopic Instrument (DESI) galaxy--quasar pairs, we map the cool circumgalactic medium (CGM) using Mg\,\textsc{ii}\,$\lambda\lambda2796,2803$ absorption with stacking technique across nearly three decades in stellar mass. 
The radial Mg\,\textsc{ii} absorption equivalent-width profile, $W_{\rm MgII}\propto D^k$, becomes progressively shallower with increasing stellar mass, following $k = 0.70\log(M_*/\mathrm{M_\odot}) - 9.12$.  Crucially, we find that the $W_{\rm MgII}$ in the CGM scales tightly with the SHMR.
The $W_{\rm Mg\,II}$ peaks at the same characteristic mass as the SHMR, coinciding with a transition of the cool CGM from super-virial to sub-virial kinematics. Low-mass halos show excess non-gravitational motions consistent with recycled accretion and feedback-driven circulation, whereas massive halos show halo-centric heating that may maintain star formation quenching.
This study bridges the gap between cool CGM and SHMR, showing that peak star formation efficiency coincides with the most abundant cool CGM reservoirs.
\end{abstract}


\section{Introduction}
\label{sec:introduction}

The relation between galaxy stellar mass and the mass of its host dark matter halo is one of the fundamental scaling relations governing galaxy formation and evolution. The stellar-to-halo mass relation (SHMR) provides a compact empirical description of how galaxies assemble their stellar mass within dark matter halos~\citep{yang03, moster10, yang12, Wechsler18}. 

The baryon-conversion efficiency, $\epsilon_*=M_*/(f_{\rm b}M_{\rm halo})$, where $f_{\rm b}=\Omega_{\rm b}/\Omega_{\rm m}$ is the cosmic baryon fraction, reaches a maximum of order $20\%$ near $M_{\rm halo}\sim10^{12}\,\mathrm{M_\odot}$ ($M_*\sim3\times10^{10}\,\mathrm{M_\odot}$) and declines toward both lower and higher masses~\citep{rodr17,kravtsov18,bilicki21,zhang22,Zacharegkas25}. The low-mass decline has motivated explanations involving baryon loss and inefficient gas-to-star conversion driven in part by stellar feedback~\citep{hopkins11,Ceverino14,Vogelsberger20,yu2025}, whereas the high-mass decline is commonly linked to the emergence of stable hot halos, virial shocks, and energetic feedback from active galactic nuclei (AGN)~\citep{Rees1977,White1978,Dekel06,bower06,croton06}. These mechanisms need not map uniquely onto the observed SHMR, but together they highlight that the efficiency peak marks a change in the balance among gas supply, recycling, heating, and removal~\citep{Wechsler18,Mitchell22}. The SHMR therefore provides a well-defined empirical characteristic scale against which changes in the baryonic components of galaxy halos can be tested, rather than a priori specifying how those components should behave.

The circumgalactic medium (CGM) provides a direct observational window on this baryon cycle because it contains gas that can be accreted, recycled, heated, or expelled before joining the stellar component~\citep{chen10cool,schaye15,tumlinson17,nelson19,hafen20,wright21,Mitchell22,Lyu-26, Ma-26}. Mg\,\textsc{ii}\,$\lambda\lambda2796,2803$ traces cool ($\sim10^4$ K), metal-enriched gas and is widely used to characterize the inner CGM~\citep{chen10absorb,Werk2013,2026enap....4..370C}. 

Previous statistical absorption and cross-correlation studies have established that Mg\,\textsc{ii} strength declines with projected distance and depends on galaxy properties including stellar mass and star-formation activity~\citep{2008ApJ...687..745C,chen10absorb,chen10cool,zhu14,2016MNRAS.455.1713H,lan18,XWChen-21,chen25}. Most of these measurements, however, have focused on the mean radial decline, covering fraction, or absorption strength at fixed radius~\citep{nielsen13b,churchill2013}. How the shape of the radial profile itself changes continuously across stellar mass remains less well constrained. This distinction is important: a mass dependence in the radial slope would indicate that the spatial organization of the cool CGM changes along the galaxy mass sequence, rather than merely reflecting a change in overall absorption amplitude. It is likewise unclear whether such structural changes occur near the same characteristic mass as changes in the absorption normalization and in the motions of the absorbing gas~\citep{nielsen2016,XWChen-21}. Establishing whether these observables vary coherently provides a direct empirical test of whether the CGM exhibits a characteristic transition comparable to the one marked by the SHMR.

DESI provides the sample size needed to trace these changes across a wide range of galaxy mass scale. We use millions of DESI Year~1 galaxy--quasar pairs~\citep{levi13,DESI2025}, comprising emission-line galaxies (ELGs), Bright Galaxy Survey galaxies (BGSs), and luminous red galaxies (LRGs) over $9<\log(M_*/\mathrm{M_\odot})<12$ and broad ranges of redshift and star-formation activity. Rather than selecting individual Mg\,\textsc{ii} systems, we median-stack quasar spectra in bins of galaxy stellar mass and projected distance $D$. We examine how the radial structure, absorption normalization, and stacked kinematics of Mg\,\textsc{ii}-bearing gas vary across the stellar-mass sequence, and whether these observables converge on a common characteristic mass scale. In this way, we seek to probe how the cool CGM is connected to its host halo across galaxy mass. Pair-by-pair $D/R_{\rm vir}$ stacks, ion-based comparisons, and within-bin redshift splits are used to test the robustness of the inferred radial trends against halo-size rescaling and population mixing. Together, these measurements provide a direct empirical test of the connection between the cool CGM and the characteristic mass scale set by its host halo.

This Letter is organized as follows. Section~\ref{sec:methods} describes the DESI sample, galaxy--quasar pair selection, spectral stacking, and absorption-line measurements.  Section~\ref{sec:radial-structure} presents the $D$-dependent structure of the cool CGM;
Section~\ref{sec:normalization} establishes the connection between the Mg\,\textsc{ii} absorption normalization and the SHMR;
and Section~\ref{sec:kinematic-transition} examines the cool-CGM kinematics from stacked absorption lines. Section~\ref{sec:discussion} discusses the broader implications and limitations, and Section~\ref{sec:summary} summarizes the main conclusions. Supporting analyses are presented in the Appendices. We adopt the Planck 2015 cosmology, with $h=0.677$, $\Omega_{\rm M}=0.309$, and $\Omega_{\Lambda}=0.691$~\citep{2016A&A...594A..13P}.

\section{Methods}
\label{sec:methods}
\subsection{DESI sample}\label{YIdata}
\begin{figure*}[t]
    \centering
    \includegraphics[width=\textwidth]{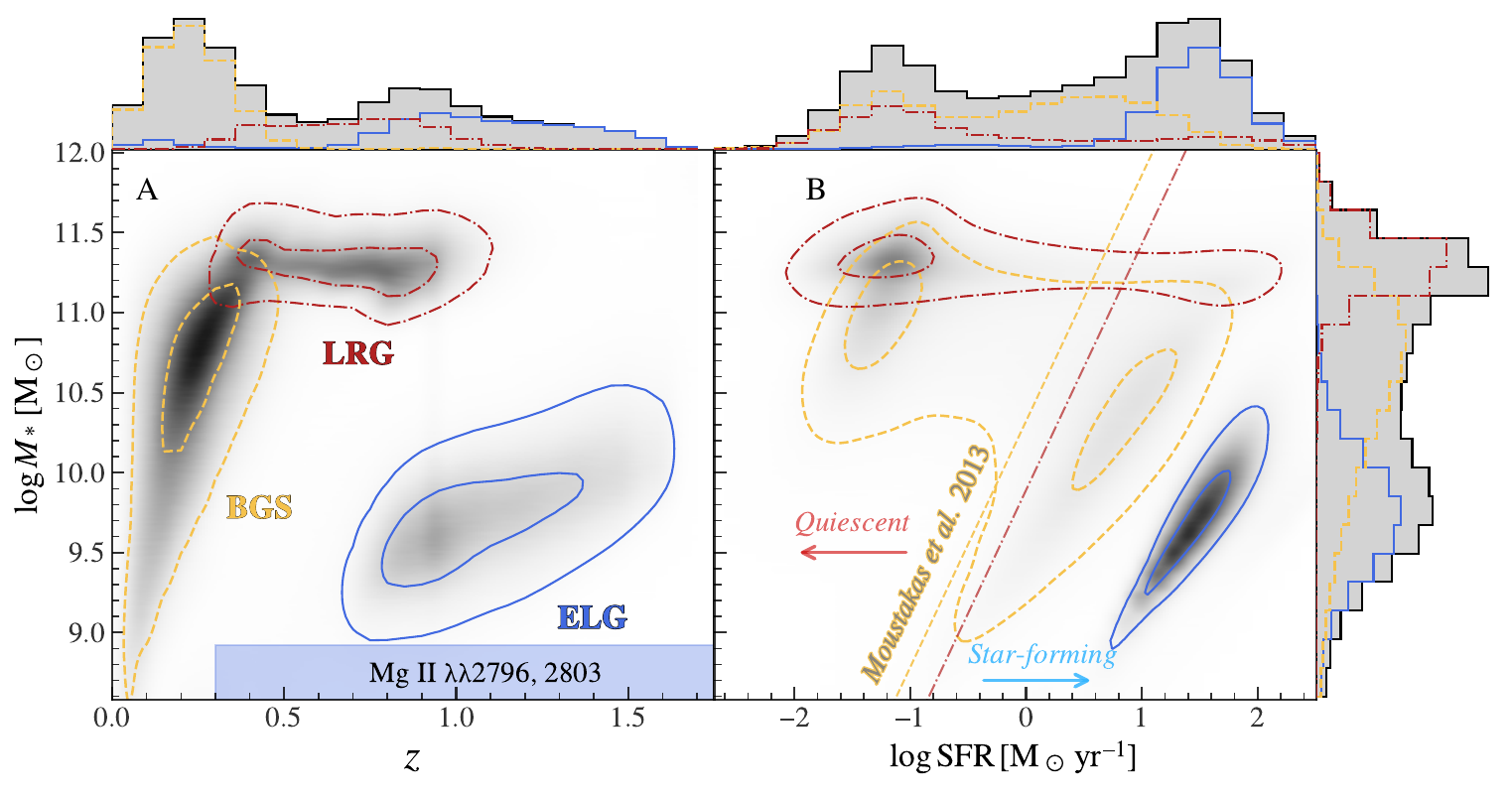}
    \caption{Stellar mass, redshift, and star-formation-rate distributions of DESI Y1 galaxies. \textbf{(A)} Stellar mass as a function of redshift, with the full distribution shown in grayscale and colored contours distinguishing the DESI galaxy classes. \textbf{(B)} Stellar mass versus SFR for the same samples. The red and orange dashed lines mark the adopted star-forming/quiescent boundaries for LRGs and BGSs~\citep{moustakas13}. Contours enclose the 68\% and 95\% regions, and the side panels show the corresponding one-dimensional distributions.}
    \label{fig:Y1}
\end{figure*}

We use spectra from the DESI Year~1 data release\footnote{\url{https://data.desi.lbl.gov/doc/releases/dr1/}}, which contains more than 18 million unique targets from Main Survey observations~\citep{DESI2025}. The relevant samples are quasars and three foreground-galaxy classes~\citep{myers23}: BGS galaxies ($0<z<0.6$, $8.5<\log M_*/\mathrm{M_{\odot}}<11.5$)~\citep{ruiz20,hahn23}, LRGs ($0.3<z<1.2$, $11.0<\log M_*/\mathrm{M_{\odot}}<12.0$)~\citep{zhou20,zhou23}, and ELGs ($0.6<z<1.6$, $8.8<\log M_*/\mathrm{M_{\odot}}<10.3$)~\citep{raichoor20,raichoor23}. Spectroscopic redshifts are obtained with \texttt{redrock}~\citep{anand24}. Stellar masses and SFRs are derived with \texttt{CIGALE}~\citep{boquien19} from $g,r,z,$ WISE 1, and WISE 2 photometry measured with \texttt{Tractor}~\citep{lang16} and compiled in the DESI Mass-EMLines catalog\footnote{\url{https://data.desi.lbl.gov/doc/releases/dr1/vac/stellar-mass-emline/}}~\citep{zou24}. The SED modeling uses a delayed star-formation history with an optional burst and Bruzual--Charlot stellar-population templates~\citep{Bruzual2003}, a Chabrier initial mass function~\citep{Chabrier03}, and Calzetti attenuation~\citep{Calzetti00}, with the spectroscopic redshift fixed to the measured value~\citep{siudek24}.

Figure~\ref{fig:Y1} summarizes the stellar-mass, redshift, and star-formation-rate distributions of the three foreground samples. The red ($\langle z \rangle = 0.62$) and orange ($\langle z \rangle = 0.36$) dashed lines in the right panel mark the adopted boundaries between star-forming and quiescent galaxies for LRGs and BGSs with redshifts greater than 0.3, respectively. The ELG sample occupies the star-forming locus, whereas most ($>80\%$) LRGs are classified as quiescent; BGS galaxies populate both regimes. The combined sample therefore spans broad, and partially overlapping, ranges of stellar mass, redshift, and star-formation activity.

\subsection{Galaxy--quasar pairs}\label{sample}
We use DESI quasars as background light sources and ELG, LRG, and BGS galaxies as foreground systems whose spectroscopic redshifts define the expected wavelengths of intervening absorption. For the Mg\,\textsc{ii} measurements, we require $z_{\rm galaxy}>0.3$ so that the Mg\,\textsc{ii}\,$\lambda\lambda2796,2803$ doublet falls within the usable DESI spectral range. This coverage requirement supersedes the lower-redshift limit used when constructing the broader parent galaxy--quasar catalogue.

We additionally require $z_{\rm galaxy}<z_{\rm QSO}-0.02(1+z_{\rm QSO})$, corresponding to a separation of approximately $6000\,{\rm km\,s^{-1}}$ in the quasar rest frame, to reduce contamination from absorption associated with the quasar host or its immediate environment~\citep{napolitano23}. This is a galaxy--quasar separation criterion, not an absorber--galaxy matching criterion. The stacking analysis does not begin with a catalogue of individually detected Mg\,\textsc{ii} absorbers, and no velocity window is imposed around a fitted absorber redshift. Instead, every pair satisfying the redshift, spectral-coverage, and projected-distance criteria is shifted to the galaxy spectroscopic rest frame and included in the median stack. Consequently, $v=0$ km$/$s in the stacked absorption profiles is defined by the galaxy spectroscopic redshift.

Within 1200~kpc, the parent sample contains 12,989,270 ELG--quasar pairs, 5,204,999 LRG--quasar pairs, and 74,505,592 BGS--quasar pairs. Their average stellar masses are $10^{9.6}\thinspace \mathrm{M_{\odot}}$, $10^{11.3}\thinspace \mathrm{M_{\odot}}$, and $10^{10.5}\thinspace \mathrm{M_{\odot}}$, respectively. Requiring the Mg\,\textsc{ii} doublet to fall within the usable DESI spectral range leaves 4,477,172 ELG--quasar pairs ($\sim$35\%), 4,588,226 LRG--quasar pairs ($\sim$88\%), and 4,030,727 BGS--quasar pairs ($\sim$5\%); supporting pair-distribution diagnostics are presented in Appendix~\ref{app:sample}.

For the primary $W_{\rm Mg\,II}$--$M_*$ analysis we use $20<D<180$ kpc. The lower limit conservatively excludes the smallest projected-separation regime, where foreground-galaxy light contamination, source deblending, and continuum-placement systematics are more difficult to control. The upper limit defines a common inner-CGM aperture while retaining sufficient stacking statistics across the stellar-mass and DESI target-class bins. Pairs at larger separations are shown for context in Figure~\ref{fig:D}, but they are not used to determine the primary $k$--$M_*$ relation.

We calculate the virial radius ($R_{\rm vir}$) following \citet{bryan98}:
\begin{equation}
R_{\rm vir} = \left( \frac{3 M_{\rm halo}}{4\pi \Delta_{\rm c}(z)\,\rho_{\rm c}(z)} \right)^{1/3},
\label{eq:Rv}
\end{equation}
For reference, 180~kpc is comparable to the characteristic virial radius inferred for galaxies near $M_*\sim3\times10^{10}\,\mathrm{M_\odot}$ at the representative redshifts of the corresponding intermediate-mass stacks. This comparison motivates the use of 180~kpc as a common inner-CGM aperture for the primary fixed-distance analysis, but 180~kpc is not assumed to be a universal virial radius. For the right panel of Figure~\ref{fig:D} only, we use the mean stellar mass and mean redshift of each fixed-$D$ stacked subsample to derive a representative $\langle R_{\rm vir}\rangle$ for display. For the ELG sample, which has the broadest redshift distribution among the three, the 1-$\sigma$ spread in redshift changes this representative virial scale by approximately $12\%$; this estimate pertains only to the bin-level display normalization. The independent halo-normalized robustness analysis instead evaluates $R_{\rm vir}$ for each galaxy at its own spectroscopic redshift and assigns each galaxy--quasar pair to a bin using its individual $D/R_{\rm vir}$ (Appendix~\ref{app:rvir}).

\subsection{Spectral stacking and absorption measurements}\label{median}\label{estimation}
To detect weak absorption features, we fit each quasar spectrum with \texttt{PyQSOFit}~\citep{guo18}\footnote{\url{https://github.com/legolason/PyQSOFit}} and normalize it by the best-fitting continuum and emission-line model. Two median-filtering passes, using 141-pixel ($\sim112.6$~\AA) and 71-pixel ($\sim56.8$~\AA) windows, suppress residual large-scale fluctuations. Five pixels ($\sim4$~\AA) centered on the absorption-line cores are masked in both passes so that the filtering does not suppress the features of interest. The processed spectra are resampled onto a common 0.2~\AA\ grid, and the median normalized flux is adopted at each pixel. Median stacking is robust to outlying sightlines and is widely used for statistical absorption measurements~\citep{bordoloi11,zhu13,zhu13ca,zhu14,bordoloi14,lan18,wu2024,ng2025}. The full set of stacked Mg\,\textsc{ii} profiles underlying the radial measurements is shown in Appendix~\ref{app:profiles}.

We use 100 bootstrap realizations to estimate the normalized-flux uncertainty at each wavelength pixel. Single- or double-Gaussian absorption models are fitted with \texttt{emcee}~\citep{foreman13}\footnote{\url{https://emcee.readthedocs.io/en/stable/index.html}} to determine the line dispersion $\sigma_{\lambda}$ and its uncertainty. We measure equivalent width by integrating the median composite over a $\pm\thinspace3$-$\sigma_{\lambda}$ interval about the best-fitting centroid and estimate its uncertainty from the root-mean-square flux uncertainty over the same interval, following \citet{chen25} and \citet{chen2025b}. All Mg\,\textsc{ii} measurements reported here use this procedure. Direct integration of the fitted Gaussian profiles gives consistent equivalent widths and leaves the results unchanged.

\subsection{Stacked absorption-line kinematics}\label{kinematics}\label{sec:kinematics-method}
In the composite spectra, the observed metal absorption features arise from ensembles of gas clouds spanning a range of velocities relative to their host galaxies. We denote the intrinsic ensemble line-of-sight width by $\sigma_v$. This quantity combines internal broadening within individual absorption systems with the distribution of bulk velocity offsets between absorbing gas and galaxies~\citep{tumlinson17,lan18}; it is therefore a statistical kinematic measure of the stacked CGM rather than a resolved velocity dispersion for individual halos. To obtain $\sigma_v$, we fit Gaussian profiles to the Mg\,\textsc{ii} $\lambda\lambda2796,2803$ doublet in the composite spectra, accounting for DESI spectral resolution and galaxy redshift uncertainties through
$\sigma_\lambda = \sqrt{\sigma_v^2 + \sigma_z^2 + \sigma_{\rm R}^2}$, where $\sigma_\lambda$ is the measured width of the stacked absorption, $\sigma_z = c\,\Delta z/(1+z)$ arises from redshift errors, and $\sigma_{\rm R} = c\Delta\lambda / [\sqrt{8\ln2}(1+z)\lambda]$ accounts for instrumental broadening~\citep{dopita03}. 

For DESI spectra, the median redshift uncertainty is $\Delta z\sim6\times10^{-5}$, and the instrumental resolution $\Delta\lambda\sim1.8$~\AA~\citep{guy2023} corresponds to a total broadening ($\sqrt{\sigma_z^2+\sigma_{\rm R}^2}$) of $\sim43$, 62 and $52~\mathrm{km\,s^{-1}}$ at 2800~\AA\ for ELGs, BGSs and LRGs, respectively. The velocity coordinate is defined relative to the galaxy spectroscopic redshift, and we obtain $\sigma_v$ by subtracting the total redshift and instrumental broadening in quadrature from the Gaussian-fitted $\sigma_\lambda$. For comparison with a characteristic halo velocity scale, we normalize this directly measured width by the line-of-sight virial velocity, $v_{\rm vir}^{\rm LOS}=v_{\rm vir}/\sqrt{3}$, where $v_{\rm vir}=\sqrt{GM_{\rm vir}/R_{\rm vir}}$~\citep{mo10}. The characteristic $M_{\rm vir}$ and $R_{\rm vir}$ are inferred from the adopted SHMR and Equation~\ref{eq:Rv}. The resulting $\sigma_v/v_{\rm vir}^{\rm LOS}$ ratio is therefore an SHMR-dependent empirical comparison, whereas $\sigma_v$ itself is measured directly from the stacked profiles after the instrumental and redshift-broadening corrections described above.

\section{Results}
\label{sec:results}
\subsection{Mass-dependent radial structure of the cool CGM}\label{sec:radial-structure}
\begin{figure*}[t]\centering
    \includegraphics[width=\textwidth]{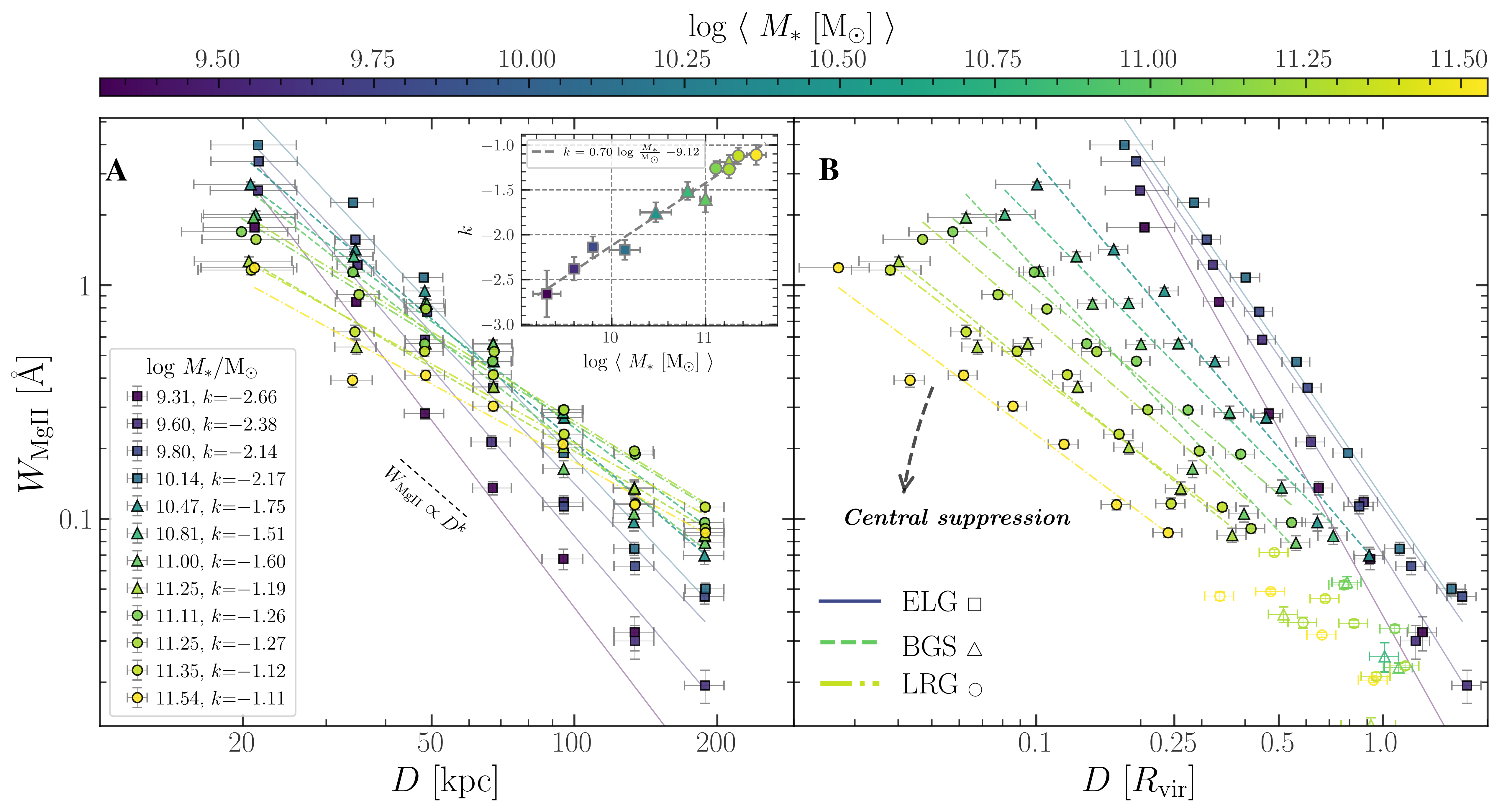}
    \caption{
$W_{\rm MgII}$ as a function of projected distance for galaxies of different stellar masses.
Different symbols represent different galaxy types: ELGs (squares\,\scalebox{0.75}{$\square$}), BGS galaxies (triangles\,\scalebox{0.85}{$\triangle$}), and LRGs (circles\,\scalebox{1.4}{$\circ$}).
\textbf{(A)} $W_{\rm MgII}$ as a function of projected distance $D$, colour-coded by $\log\langle M_*/\mathrm{M_\odot}\rangle$. The best-fit power-law profiles within 180~kpc are shown for ELGs (solid), BGS galaxies (dashed), and LRGs (dotted). The inset shows the corresponding $k$--$\log M_*$ relation. Both the inset and the displayed profiles are based on stacks constructed in fixed physical-distance bins.
\textbf{(B)} the same fixed-$D$ measurements displayed against $D/R_{\rm vir}$ using the representative $\langle R_{\rm vir}\rangle$ of each stellar-mass bin. A separate pair-by-pair $D/R_{\rm vir}$ re-stacking analysis is described in Appendix~\ref{app:rvir}. Horizontal error bars show the 1-$\sigma$ dispersion of projected distances contributing to each stack, while vertical error bars show the 1-$\sigma$ uncertainty in $W_{\rm MgII}$. The 180~kpc boundary is used as a common inner-CGM aperture rather than a universal virial radius. Measurements beyond 180~kpc and out to $\sim1\,R_{\rm vir}$ are shown as open symbols and broadly follow the extrapolation of the inner power-law fits.
}
    \label{fig:D}
\end{figure*}

We divide the galaxies into 12 stellar-mass populations, four for each DESI target class, and stack the associated quasar spectra in seven projected-distance bins (Section~\ref{sec:methods}). This provides a two-dimensional sampling of $W_{\rm MgII}$ in the $M_*\!-D$ plane (Figure~\ref{fig:D}). The measured $W_{\rm MgII}(D)$ profiles and the fitted $k$--$M_*$ relation are two representations of the same observational result. Within 180~kpc, the profiles are well described by power laws, $W_{\rm MgII}\propto D^k$, consistent with earlier statistical Mg\,\textsc{ii} studies~\citep{2008ApJ...687..745C,chen10absorb,chen10cool,zhu14,lan18,XWChen-21,chen25,kong26}. The fitted slope becomes progressively less negative with increasing $M_*$ and is described by $k=0.70\,\log(M_*/\mathrm{M_\odot})-9.12$ over the sampled range. Thus, lower-mass systems have more centrally concentrated Mg\,\textsc{ii} absorption profiles, whereas those around more massive systems decline more gradually with projected distance.

Importantly, the change in slope forms a continuous sequence across the sampled stellar-mass range rather than a simple offset among the ELG, BGS, and LRG target classes. The profiles are therefore not related by a common radial shape with only a mass-dependent normalization: their shapes themselves change systematically with $M_*$. Measurements beyond 180~kpc, although excluded from the primary power-law fits, broadly continue the corresponding inner trends out to halo-scale radii. 

The mass-dependent change in profile shape is most apparent in the inner CGM. At the high-mass end, the decline in Mg\,\textsc{ii} absorption relative to the intermediate-mass maximum is strongest at small projected radii and becomes weaker toward larger, halo-scale radii (Figures~\ref{fig:D} and~\ref{fig:normalization}). We refer to this differential behaviour descriptively as a central suppression of Mg\,\textsc{ii} absorption. Heating by a virialized hot halo or AGN feedback offers a plausible explanation for reducing the central contribution of Mg\,\textsc{ii}--bearing gas~\citep{Dekel06,fabian12}. The changing value of $k$ is therefore best interpreted as a mass-dependent reshaping of the radial Mg\,\textsc{ii} absorption profile, rather than simply a change in its overall normalization.

We perform several tests to assess the robustness of this mass-dependent radial trend. The mass dependence is retained when each galaxy--quasar pair is assigned to a radial bin using its individual $D/R_{\rm vir}$ before stacking (Appendix~\ref{app:rvir}), showing that the $k$--$M_*$ relation is not produced solely by applying the same physical-distance bins to halos of different characteristic size. The same qualitative slope--mass behaviour is recovered with Fe\,\textsc{ii}\,$\lambda2600$, and the low- and high-redshift halves of each Mg\,\textsc{ii} mass bin agree within their uncertainties (Appendix~\ref{app:krobust}). These tests establish the $k$--$M_*$ relation as a robust mass-dependent trend rather than a consequence of halo-size or redshift effects, suggesting a continuous connection between the structure of the cool CGM and the overall mass scale of the galaxy--halo system.

\subsection{The Mg\,\textsc{ii} normalization and its empirical connection to the SHMR}\label{sec:normalization}
\begin{figure*}[t]
    \centering
    \includegraphics[width=\textwidth]{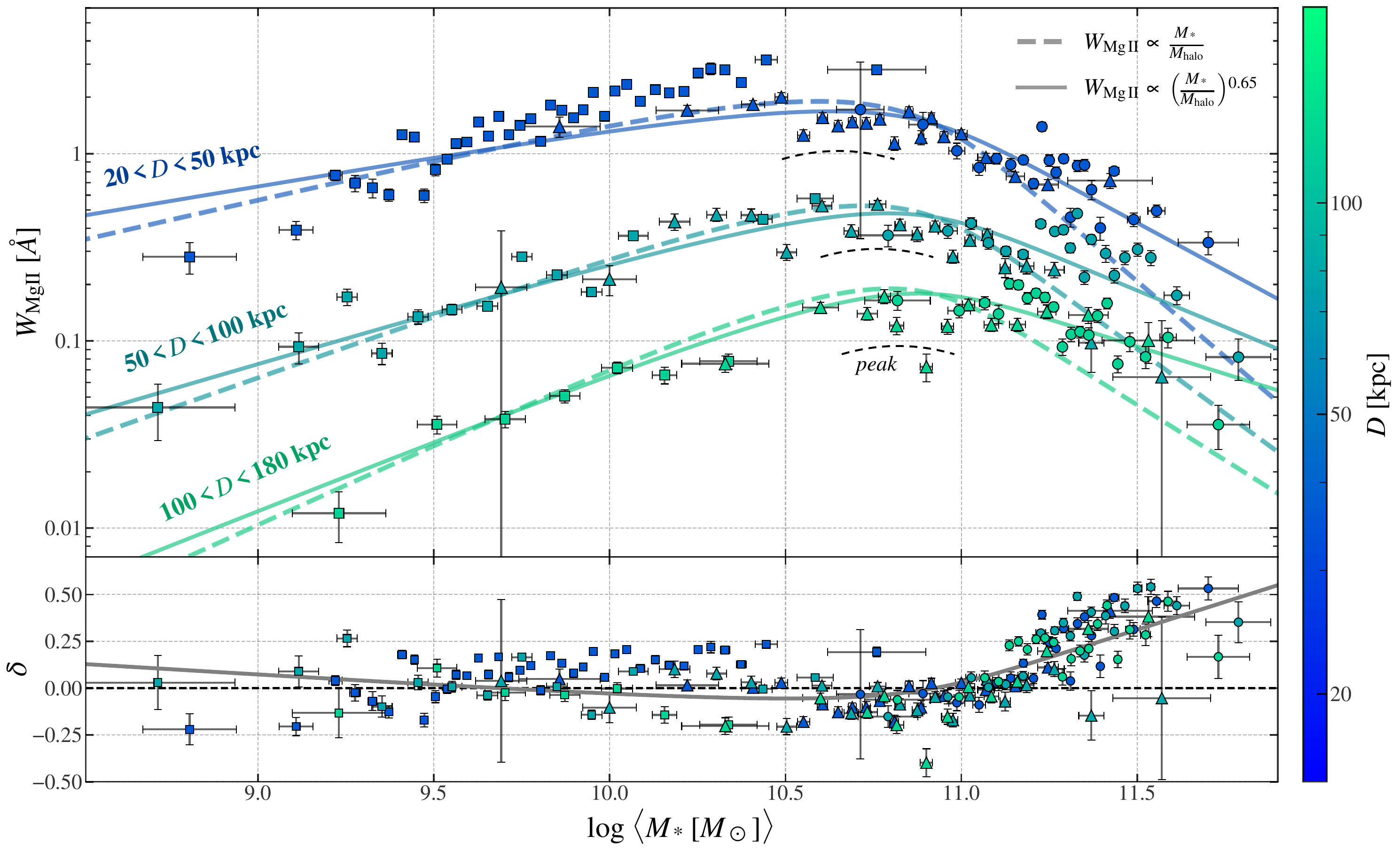}
    \caption{Mass dependence of the Mg\,\textsc{ii} absorption normalization. The upper panel shows $W_{\rm MgII}$ as a function of $\log\langle M_*\rangle$ in three projected-distance intervals; the mean distances are encoded by the colour bar. Squares, triangles, and circles denote ELGs, BGSs, and LRGs, respectively. Dashed and solid curves show the two empirical scalings discussed in Section~\ref{sec:normalization}, $W_{\rm MgII}\propto M_*/M_{\rm halo}$ and $W_{\rm MgII}\propto(M_*/M_{\rm halo})^{0.65}$; the short black dashed curves indicate the maxima of the fiducial relations. The lower panel shows residuals from the fiducial model, $\delta=\log W_{\rm MgII}-\log W_{\rm MgII}^{\rm model}$; the grey curve gives the residual expected for the generalized scaling relative to the fiducial model. Horizontal error bars show the 1-$\sigma$ stellar-mass dispersion and vertical error bars show the 1-$\sigma$ uncertainty in $W_{\rm MgII}$.}
    \label{fig:normalization}
\end{figure*}

\begin{figure*}[t]
    \centering
    \includegraphics[width=\textwidth]{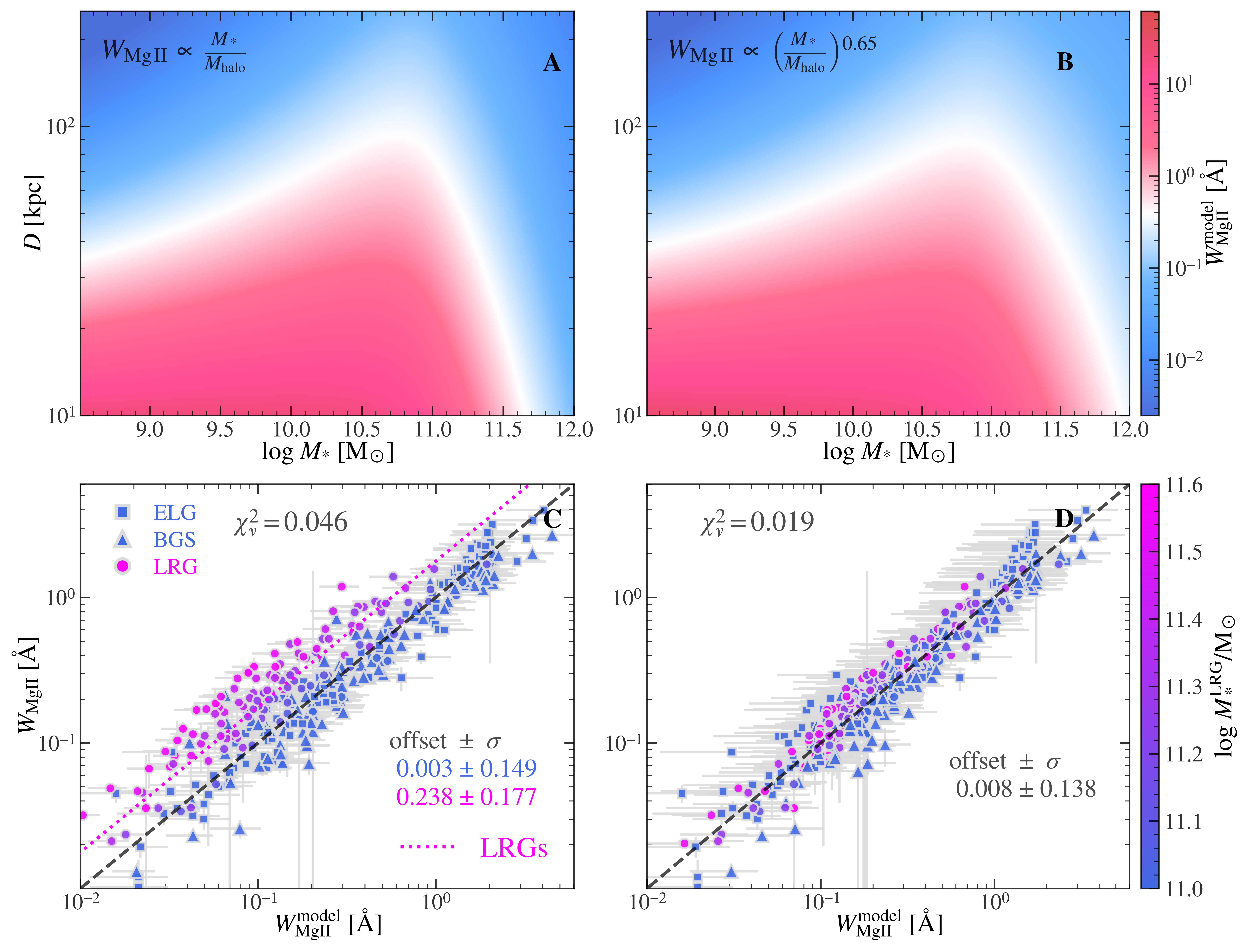}
    \caption{Empirical Mg\,\textsc{ii} absorption models in the stellar-mass--projected-distance plane. \textbf{(A)} Two-dimensional prediction of the fiducial SHMR-based model, $W^{\rm model}_{\rm MgII}\propto M_*/M_{\rm halo}$ (Equation~\ref{eq:QR}), fitted to ELGs and BGSs. \textbf{(B)} Corresponding prediction for the generalized scaling, $W^{\rm model'}_{\rm MgII}\propto(M_*/M_{\rm halo})^{0.65}$, fitted to all three galaxy classes. \textbf{(C)} Comparison of the fiducial model with the stacked measurements. \textbf{(D)} Comparison of the generalized model with the same measurements. In (C) and (D), squares, triangles, and circles denote ELGs, BGSs, and LRGs, respectively; grey error bars show measurement uncertainties, and the colour bar indicates stellar mass. Text labels give the mean offset and 1-$\sigma$ scatter for ELGs+BGSs (blue), LRGs (pink), and all galaxies (black); the uncertainty in $W_{\rm MgII}^{\rm model}$ includes the 1-$\sigma$ scatter in $M_*/M_{\rm halo}$~\citep{moster18}. The largest systematic residuals occur for the most massive LRGs.}
    \label{fig:model}
\end{figure*}

We next ask how the overall absorption normalization varies along the same stellar-mass sequence. We separate the radial and mass dependencies as
\[
W_{\rm MgII}=f(M_*)\left(\frac{D}{D_0}\right)^{k(M_*)},
\]
where $k(M_*)$ specifies the radial profile shape and $f(M_*)$ denotes its mass-dependent normalization at a fixed reference scale $D_0$. The mass dependence of this normalization can be traced directly by comparing $W_{\rm MgII}$ at fixed projected distance. We therefore measure $W_{\rm MgII}$ in the 20--50, 50--100, and 100--180~kpc intervals using finer stellar-mass sampling than in the radial-profile fits of Figure~\ref{fig:D}. 
The finer sampling resolves the mass dependence of the absorption strength, whereas the broader bins in Figure~\ref{fig:D} provide the signal-to-noise required to determine a radial slope in each mass interval. In all three distance ranges, $W_{\rm MgII}$ rises toward intermediate stellar mass and then declines at the high-mass end, producing a broad maximum near $M_*\sim3\times10^{10}\,\mathrm{M_\odot}$ (Figure~\ref{fig:normalization}). The turnover is established directly from the stacked measurements and is retained when the analysis is restricted to $0.6<z<1.1$, where the ELG and LRG samples overlap substantially in redshift (Appendix~\ref{app:redshift-ssfr}).

An empirical connection between the SHMR and the cool CGM is physically plausible because both are shaped by the balance among baryon accretion, gravitational confinement, stellar feedback, and halo heating~\citep{tumlinson17,Mitchell22}. At lower halo masses, feedback can efficiently expel or recycle cool gas, while near $M_{\rm halo}\sim10^{12}\,\mathrm{M_\odot}$, deeper potentials and the emergence of stable hot halos can alter the retention, ionization, and survival of low-ionization gas~\citep{Dekel06,Muratov-15,Mitchell22, yu2025}. The same halo-scale processes that regulate stellar-mass assembly may therefore imprint a characteristic mass scale on the Mg\,\textsc{ii}-bearing CGM.

Guided by this physical expectation and by the proximity of the observed broad maximum to the characteristic stellar mass at which the SHMR peaks, we use the \citet{moster18} relation as a compact empirical template for the normalization (Appendix~\ref{app:halo}). We write $f(M_*)=f_0\,M_*/M_{\rm halo}$.

We fit the ELG and BGS measurements with $f_0$ and $D_0$ as free parameters. The resulting relation is
\begin{equation}
\begin{array}{rcl}
W^{\rm model}_{\rm MgII}/\text{\AA} & \simeq &
22.4\,\left(\frac{D}{63.4\,\mathrm{kpc}}\right)^k
\frac{M_*}{M_{\rm halo}},\\
k & = & 0.70\,\log(M_*/\mathrm{M}_{\odot})-9.12.
\end{array}
\label{eq:QR}
\end{equation}
Figures~\ref{fig:normalization} and~\ref{fig:model} compare this empirical scaling with the DESI Year~1 stacks. The fiducial model recovers the location of the broad maximum at different projected distances and describes the ELG and BGS measurements with a scatter of $\sim0.15$ dex. The two-dimensional comparison in Figure~\ref{fig:model} shows that this agreement extends across much of the sampled $M_*\!-D$ plane. This empirical scaling also provides a useful first-order description of previous Mg\,\textsc{ii} measurements~\citep[e.g.,][]{bordoloi11,nielsen13,nielsen13b,dutta20} spanning a broad range of galaxy populations and projected distances (Appendix~\ref{app:previous}).

The largest systematic departures occur at the high-mass end. The simplest $M_*/M_{\rm halo}$ scaling underpredicts the most massive LRG measurements by an average of $\sim0.24$ dex, with the residual increasing toward the highest stellar masses, as shown in the left panels of Figure~\ref{fig:model}. This offset is qualitatively consistent with weak-lensing results showing that red, quenched galaxies can occupy more massive halos than star-forming galaxies at fixed $M_*$~\citep{Velander2014,Mandelbaum16}. Differences in star-formation history, environment, ionization state, and residual AGN activity may also contribute to the LRG departure~\citep{fabian12,tumlinson17,2026enap....4..370C}. A more flexible mass dependence, $f(M_*)=f_0(M_*/M_{\rm halo})^n$, gives $W^{\rm model'}_{\rm MgII}/\text{\AA}\simeq10.2\,(D/46.3\,\mathrm{kpc})^k(M_*/M_{\rm halo})^{0.65}$ and reduces the overall scatter to $\sim0.14$ dex (the right panels of Fig~\ref{fig:model}), although a systematic high-mass residual remains. Despite their different mass dependences, both models show that the absorption normalization identifies a characteristic mass scale close to the peak of the SHMR.

\subsection{A kinematic transition near the same characteristic mass}\label{sec:kinematic-transition}
\begin{figure*}[t]
    \centering
    \includegraphics[width=\textwidth]{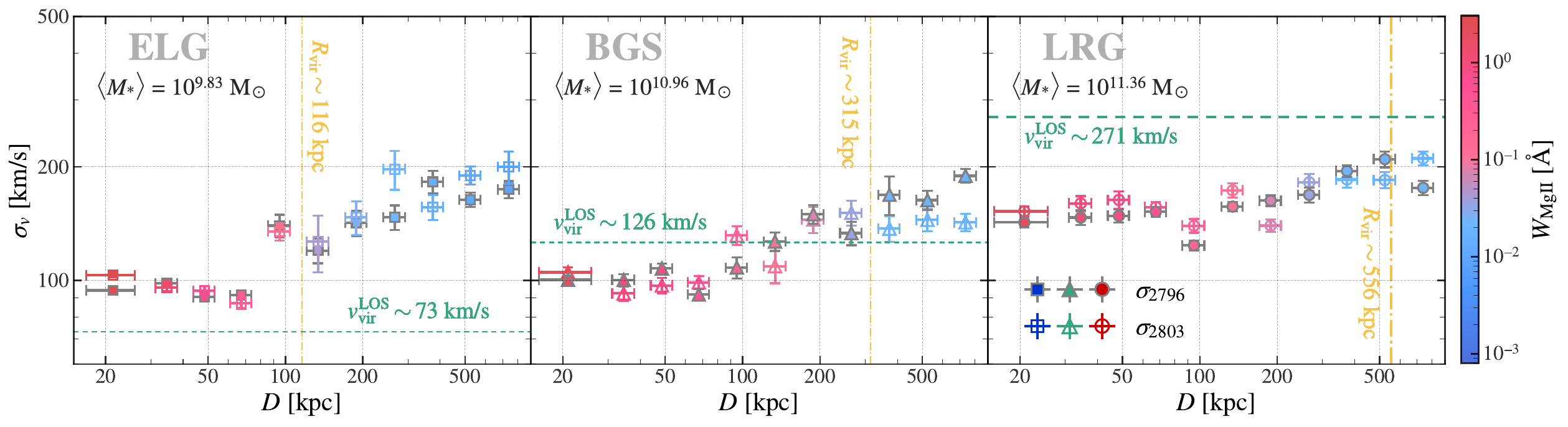}
    \par\vspace{0.5em}
    \includegraphics[width=\textwidth]{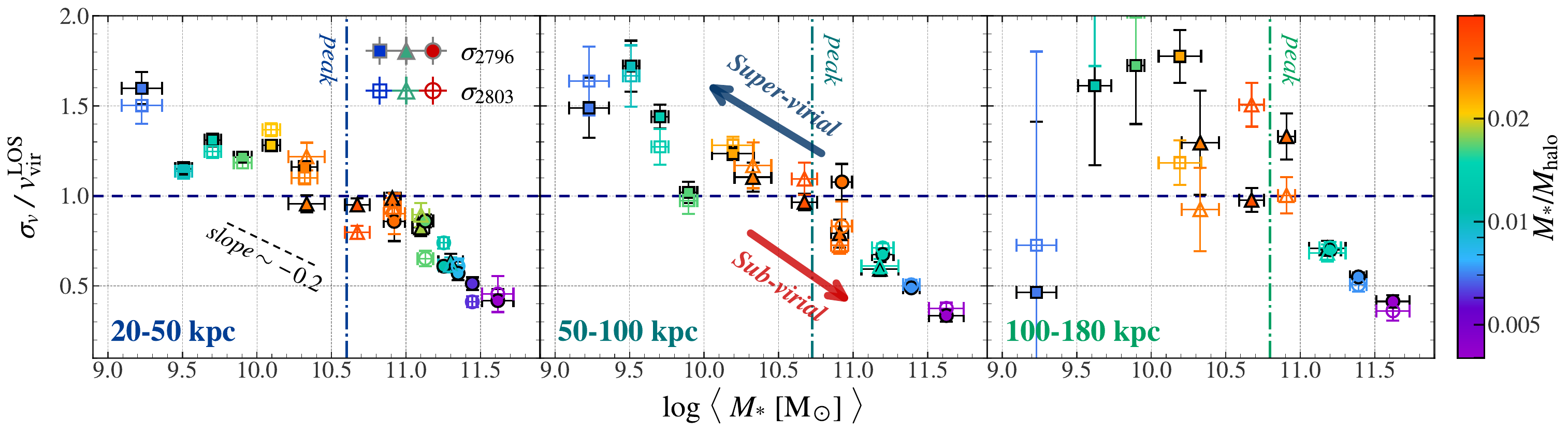}
    \caption{Radial profiles and halo-normalized mass dependence of the stacked Mg\,\textsc{ii} velocity width. The upper row shows the corrected line-of-sight width, $\sigma_v$, as a function of projected distance for representative ELG, BGS, and LRG stacks with $\langle M_*\rangle=10^{9.83}$, $10^{10.96}$, and $10^{11.36}\,\mathrm{M_\odot}$, respectively. Filled and open symbols show Mg\,\textsc{ii} $\lambda2796$ and $\lambda2803$; colours encode $W_{\rm MgII}$. Vertical dot-dashed lines mark $R_{\rm vir}\simeq116$, 315, and 556~kpc, and horizontal dashed lines mark $v_{\rm vir}^{\rm LOS}\simeq73$, 126, and $271~\mathrm{km\,s^{-1}}$. The lower row shows $\sigma_v/v_{\rm vir}^{\rm LOS}$ as a function of stellar mass in the 20--50, 50--100, and 100--180~kpc intervals. Squares, triangles, and circles denote ELGs, BGSs, and LRGs; filled and open symbols show the two doublet transitions, and colours encode $M_*/M_{\rm halo}$. The horizontal dashed line marks unity, while vertical dot-dashed lines mark the peak mass of the fiducial absorption model in each distance interval. The ratio crosses unity near $\log(M_*/\mathrm{M_\odot})\simeq10.5$. Error bars show 1-$\sigma$ measurement uncertainties and the dispersion of the contributing projected-distance or stellar-mass distributions.}
    \label{fig:kinematic-transition}
\end{figure*}

The stacked Mg\,\textsc{ii} line widths provide a third observational dimension along the same stellar-mass sequence. The upper row of Figure~\ref{fig:kinematic-transition} shows the directly measured radial profiles of the corrected ensemble width, $\sigma_v(D)$, for representative ELG, BGS, and LRG stacks. The instrumental and redshift-broadening corrections are described in Section~\ref{sec:kinematics-method}; here we focus on the resulting kinematic trends.

For the representative ELG stack, much of the measured $\sigma_v(D)$ profile lies above its adopted $v_{\rm vir}^{\rm LOS}$ expectation, with the excess being most apparent at relatively small projected radii. The BGS profile approaches that scale, whereas the LRG profile is predominantly below it. The Mg\,\textsc{ii} $\lambda2796$ and $\lambda2803$ measurements show the same qualitative radial behaviour in all three populations. The vertical $R_{\rm vir}$ and horizontal $v_{\rm vir}^{\rm LOS}$ guides are characteristic halo scales inferred from the fiducial SHMR, not direct dynamical measurements of the individual systems.

Combining all stellar-mass bins in the three projected-distance intervals reveals a mass-dependent transition in the characteristic kinematic state of the cool CGM (lower row of Figure~\ref{fig:kinematic-transition}). The ratio $\sigma_v/v_{\rm vir}^{\rm LOS}$ decreases systematically with stellar mass: at low mass, the Mg\,\textsc{ii}-bearing gas is predominantly super-virial relative to the adopted line-of-sight expectation, particularly in the inner CGM, indicating that the observed motions are not set by halo gravity alone and likely include substantial non-equilibrium contributions from feedback-driven circulation, recycled accretion, or other bulk motions. As stellar mass increases, the ratio declines continuously and the cool CGM moves progressively into the sub-virial regime~\citep{2016MNRAS.455.1713H}. This kinematic transition is qualitatively consistent with the independent DESI results of \citet{Yesuf2026}, who find gravitationally bound cool-gas outflows to dominate in star-forming galaxies, while inflows are more commonly detected in quiescent systems. Their picture of feedback-driven recycling in star-forming galaxies and slowly cooling enriched gas in quiescent systems provides a complementary physical interpretation of this transition, suggesting an evolution from feedback-dominated circulation toward more gravitationally confined and potentially settling cool gas in the massive, quenched regime.


Notably, this kinematic transition from super- to sub-virial occurs at the same characteristic mass scale ($M_*\sim3\times10^{10}\,\mathrm{M_\odot}$) where the radial structure of the cool CGM changes systematically and the Mg\,\textsc{ii} absorption normalization reaches its maximum. This concurrence suggests that the changing radial profile and the kinematic transition are not independent phenomena, but may represent two observational manifestations of the same mass-dependent transformation of the cool CGM: one in its spatial distribution and the other in its dynamical state. These results show that the radial structure, absorption normalization, and stacked kinematics all identify a common empirical change near the mass of peak galaxy-formation efficiency. This concurrence, rather than exact agreement with any single SHMR prescription, is the principal empirical result.

\section{Discussion}
\label{sec:discussion}
The DESI Year~1 stacks place three aspects of the cool CGM on a common stellar-mass sequence: the radial profile shape, the absorption normalization, and the characteristic stacked line width. Previous statistical studies established the radial decline of Mg\,\textsc{ii} absorption and its dependence on galaxy properties~\citep{chen10cool,zhu14,lan18,XWChen-21,chen25}; the present measurements show that these three observables change coherently near $M_*\sim3\times10^{10}\,\mathrm{M_\odot}$. The proximity of this scale to the peak of the SHMR~\citep{Wechsler18,moster18} makes galaxy-formation efficiency a useful empirical organizing framework, but the agreement should not be interpreted as evidence that the SHMR uniquely determines the instantaneous Mg\,\textsc{ii}-bearing CGM.

The high-mass behaviour further suggests that this transition involves not only the overall strength of the cool CGM, but also its radial and phase structure (Figure~\ref{fig:D}). The suppression of Mg\,\textsc{ii} absorption is strongest at small projected radii, qualitatively consistent with enhanced heating or ionization in the inner CGM from a virialized hot halo or AGN feedback, which can reduce the survival or covering fraction of Mg\,\textsc{ii}-bearing gas. At the same time, the most massive LRGs show systematically stronger Mg\,\textsc{ii} absorption than predicted by the simplest $M_*/M_{\rm halo}$ scaling. Weak-lensing studies indicate that red, quenched galaxies can occupy more massive halos than star-forming galaxies at fixed stellar mass~\citep{Velander2014,Mandelbaum16}; such systems may therefore host larger total baryonic reservoirs, potentially allowing a more substantial cool component (Figure~\ref{fig:model}). In this picture, the high-mass CGM is not simply depleted, but is reorganized in radius and phase, with central suppression coexisting with substantial Mg\,\textsc{ii} absorption on larger halo scales.

\section{Summary}
\label{sec:summary}
Using median stacks of millions of DESI Year~1 galaxy--quasar pairs, we measured the radial structure, absorption normalization, and characteristic stacked kinematics of Mg\,\textsc{ii}-bearing cool CGM gas across nearly three decades in stellar mass. The principal results are:

\textit{Mass-dependent radial structure.} Within 180~kpc, the Mg\,\textsc{ii} equivalent-width profiles are well described by $W_{\rm MgII}\propto D^k$, with $k=0.70\,\log(M_*/\mathrm{M_\odot})-9.12$ over the sampled mass range (Figure~\ref{fig:D}). Lower-mass systems have more centrally concentrated profiles, whereas the profiles become progressively shallower toward higher stellar mass. The trend persists when galaxy--quasar pairs are re-binned using their individual $D/R_{\rm vir}$, and similar slope--mass behaviour is recovered with Fe\,\textsc{ii} and in within-bin redshift tests (Appendix~\ref{app:robustness}). At the high-mass end, the reduction in Mg\,\textsc{ii} absorption is strongest at small projected radii, indicating a mass-dependent reshaping of the inner-CGM absorption profile rather than a simple halo-size rescaling.

\textit{Absorption normalization and its empirical connection to the SHMR.} At fixed projected distance, $W_{\rm MgII}$ rises toward intermediate stellar mass, reaches a broad maximum near $M_*\sim3\times10^{10}\,\mathrm{M_\odot}$, and declines toward the most massive systems (Figure~\ref{fig:normalization}). This turnover is measured directly from the stacks and lies close to the characteristic stellar mass at which the SHMR peaks. An empirical $M_*/M_{\rm halo}$ scaling describes the ELG and BGS measurements with a scatter of $\sim0.15$ dex, but underpredicts the most massive LRGs by an average of $\sim0.24$ dex. A generalized $(M_*/M_{\rm halo})^{0.65}$ scaling reduces the overall scatter to $\sim0.14$ dex, while leaving a systematic high-mass residual (Figure~\ref{fig:model}). We therefore treat the SHMR as a useful organizing description of the normalization, not as a unique physical determinant of the cool-CGM reservoir.

\textit{Stacked cool-CGM kinematics.} The directly measured $\sigma_v(D)$ profiles evolve from widths that are commonly above the adopted line-of-sight virial expectation in representative ELG stacks to widths that lie predominantly below it in representative LRG stacks, with consistent behaviour in both Mg\,\textsc{ii} doublet transitions. Across the full stellar-mass sequence, $\sigma_v/v_{\rm vir}^{\rm LOS}$ decreases with mass and crosses unity near $\log(M_*/\mathrm{M_\odot})\simeq10.5$ (Figure~\ref{fig:kinematic-transition}). Its concurrence with the absorption turnover suggests that the halo transition near peak galaxy-formation efficiency is also imprinted on the ensemble dynamics of the Mg\,\textsc{ii}-bearing CGM.

Taken together, the changing radial profile shape, the turnover in absorption normalization, and the kinematic crossing identify an empirical cool-CGM transition near the stellar mass of peak galaxy-formation efficiency. The concurrence of these observables is the central result; its physical origin need not be uniquely specified by the SHMR but provides a compact benchmark for models of baryon cycling, halo heating, and the survival of cool gas across the galaxy mass sequence.

\section*{Acknowledgements}
This research used data obtained with the Dark Energy Spectroscopic Instrument (DESI). DESI construction and operations are managed by Lawrence Berkeley National Laboratory. This material is based upon work supported by the U.S. Department of Energy, Office of Science, Office of High Energy Physics, under Contract No. DE--AC02--05CH11231, and by the National Energy Research Scientific Computing Center, a DOE Office of Science User Facility under the same contract. Additional support for DESI was provided by the U.S. National Science Foundation (NSF), Division of Astronomical Sciences under Contract No. AST-0950945 to the NSF's National Optical-Infrared Astronomy Research Laboratory; the Science and Technology Facilities Council of the United Kingdom; the Gordon and Betty Moore Foundation; the Heising-Simons Foundation; the French Alternative Energies and Atomic Energy Commission; the National Council of Humanities, Science and Technology of Mexico; the Ministry of Science and Innovation of Spain; and the DESI Member Institutions listed at \url{https://www.desi.lbl.gov/collaborating-institutions}. The DESI collaboration is honored to be permitted to conduct scientific research on I'oligam Du'ag (Kitt Peak), a mountain with particular significance to the Tohono O'odham Nation.

We thank Yongda Zhu for helpful discussions and contributions during the early development of this work.

This work was supported by the National Natural Science Foundation of China (grant Nos. 12473008, 12192224, and 12595312). H.W. was supported by the New Cornerstone Science Foundation through the XPLORER PRIZE. L.C.H. was supported by the National Natural Science Foundation of China (grant No. 12233001) and the China Manned Space Program (grant No. CMS-CSST-2025-A09). H.Z. was supported by the National Natural Science Foundation of China (grant Nos. 12120101003 and 12373010), the National Astronomical Observatories, Chinese Academy of Sciences (project No. E5ZQ7801), and the China Manned Space Program (grant No. CMS-CSST-2025-A06).

\appendix
\setcounter{figure}{0}
\setcounter{equation}{0}
\renewcommand{\thefigure}{\Alph{section}\arabic{figure}}
\renewcommand{\theHfigure}{appendix.\Alph{section}.\arabic{figure}}
\renewcommand{\theequation}{\Alph{section}\arabic{equation}}
\renewcommand{\theHequation}{appendix.\Alph{section}.\arabic{equation}}

\section{Model and Measurement Diagnostics}\label{app:diagnostics}
\subsection{SHMR and population dependence}\label{app:halo}
\begin{figure*}[t]
    \centering
    \includegraphics[width=\textwidth]{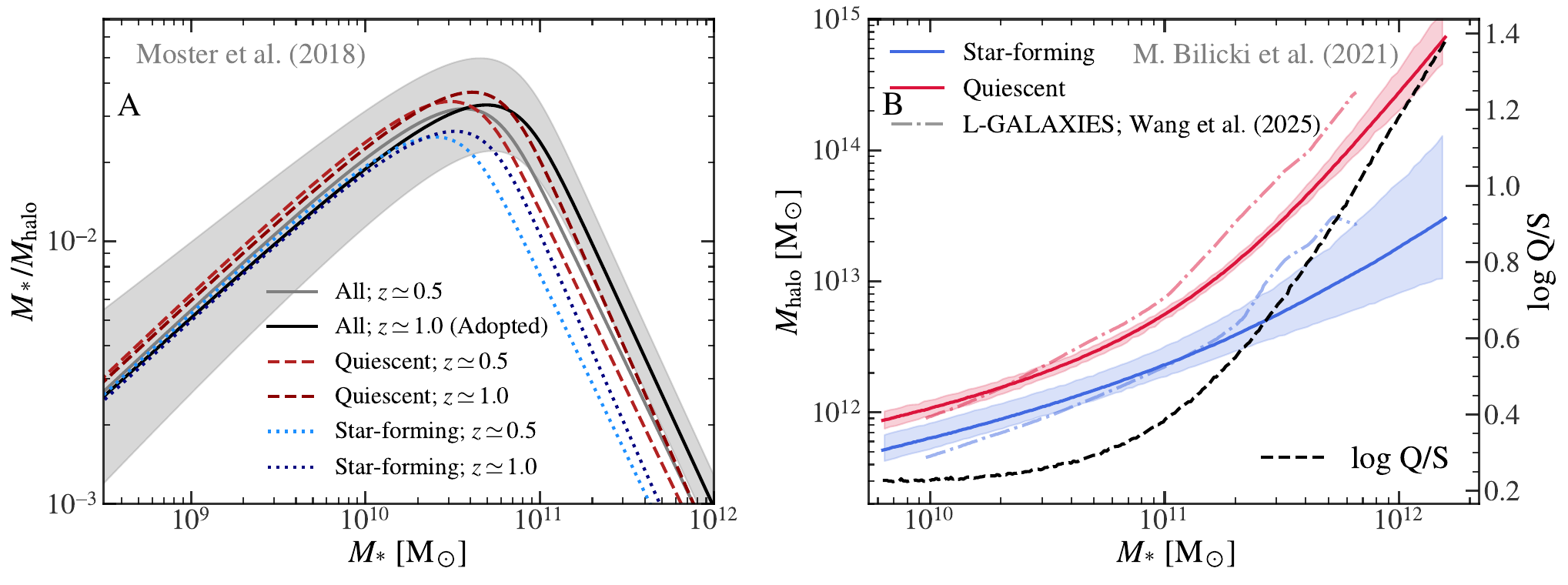}
    \caption{Stellar-to-halo mass relation and halo mass growth for different galaxy populations. 
\textbf{(A)} Stellar-to-halo mass ratio $M_*/M_{\rm halo}$ as a function of stellar mass $M_*$ at $z\approx0.5$ and $z\approx1.0$. 
The black curve and grey band show the relation and $1\,\sigma$ uncertainty for the full galaxy sample at $z\sim1$ from \citet{moster18}. 
Red and blue curves denote quenched and star-forming centrals, respectively, at the two redshifts.  
\textbf{(B)} Median halo mass $M_{\rm halo}$ versus $M_*$ at low redshift from \citet{bilicki21}, for star-forming (blue) and quiescent (red) centrals; shaded regions are their $1\,\sigma$ confidence intervals. The dot-dashed lines show the same quantities predicted by the L-GALAXIES semi-analytic model~\citep{wang2025}.
The black dashed curve indicates the logarithm of the quiescent-to-star-forming halo-mass ratio, $\log(Q/S)$, derived from \citet{bilicki21}. }
    \label{fig:moster}
\end{figure*}

We use the SHMR as an empirical mapping between stellar mass and a characteristic halo mass, and as a reference description of the stellar-mass scale at which baryon-conversion efficiency peaks~\citep{kravtsov18,legrand19,girelli20}. The general SHMR may be written as a function of halo mass and redshift~\citep{moster13}:
\begin{equation}
\frac{M_*}{M_{\rm halo}}(M_{\rm halo}, z) = 2\,f_b\,\epsilon_{\mathrm{N}} \left[ \left( \frac{M_{\rm halo}}{M_1} \right)^{-\beta} + \left( \frac{M_{\rm halo}}{M_1} \right)^{\eta} \right]^{-1}.
\label{eq:SHMR}
\end{equation}
Here $f_b=0.157$ is the universal baryon fraction. For the present analysis we adopt a single fiducial parameter set, $\epsilon_{\mathrm N}=0.19$, $M_1=10^{11.98}\,\mathrm{M_\odot}$, $\beta=1.53$, and $\eta=0.59$, corresponding to the $z\simeq1$ relation reported by \citet{moster18}. We use this relation consistently to define characteristic $M_{\rm halo}$ and $R_{\rm vir}$ values across the sample. The characteristic form and turnover scale of the SHMR evolve only moderately over the redshift interval relevant here~\citep{moster13,girelli20,shuntov22}; nevertheless, the inferred halo quantities should be regarded as characteristic scales rather than precision halo-mass estimates for individual galaxies.

The SHMR exhibits population dependence, with quiescent galaxies tending to reside in more massive halos at fixed stellar mass~\citep{more11,wojtak13,Velander2014,Behroozi19,Posti21}, as illustrated in the right panel of Figure~\ref{fig:moster}. At fixed $M_*$, quiescent centrals can inhabit dark-matter halos that are $\sim0.3$--$0.5$ dex more massive than those of star-forming centrals~\citep{bilicki21}. Such offsets can reflect differences in assembly history and quenching, and they motivate treating a single SHMR as a characteristic rather than population-specific halo-mass mapping. In the \texttt{L-GALAXIES} semi-analytic model, an early phase of rapid star formation is followed in a subset of systems by black-hole feedback that suppresses subsequent in-situ stellar growth, whereas star-forming centrals continue to grow at later epochs~\citep{wang2025}.

\subsection{Sample diagnostics}\label{app:sample}
\begin{figure*}[t]
    \centering
    \includegraphics[width=\textwidth]{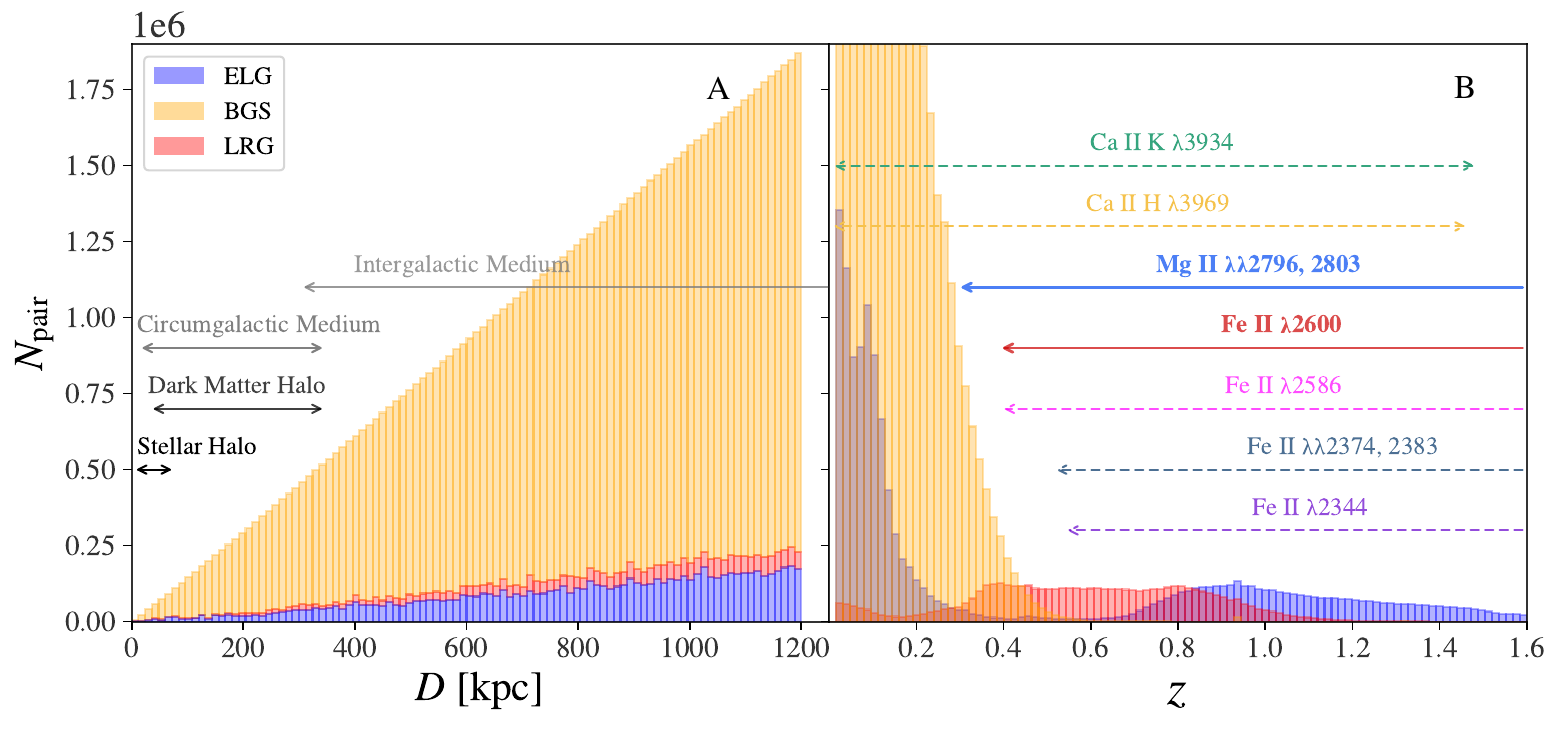}
    \caption{Distribution of galaxy--quasar pairs in projected distance and redshift. Histograms show the number of galaxy--quasar pairs (\textit{N\textsubscript{pair}}) as a function of projected distance $D$ in (A) and redshift $z$ in (B) for three galaxy populations in the DESI Y1 sample: ELGs (blue), LRGs (red), and BGSs (orange). Grey horizontal arrows reproduce the schematic physical-scale guides shown in the left panel and should not be interpreted as sharp boundaries between stellar halos, dark matter halos, the CGM, and the IGM. In the right panel, horizontal colored arrows mark the redshift ranges over which various absorption features---such as Ca\,\textsc{ii} H and K, Mg\,\textsc{ii}, and Fe\,\textsc{ii} transitions---are covered by the DESI spectral range. The two lines adopted in this work, Mg\,\textsc{ii}\,$\lambda\lambda$2796, 2803 and Fe\,\textsc{ii}\,$\lambda$2600, are highlighted in bold.}
    \label{fig:pair}
\end{figure*}

Figure~\ref{fig:pair} summarizes the galaxy--quasar pair distribution underlying the stacked measurements and provides context for the selection described in Section~\ref{sample}. The projected-distance distribution extends well beyond the primary $20<D<180$~kpc interval used for the main $W_{\rm MgII}$--$M_*$ analysis, allowing the larger-$D$ measurements in Figure~\ref{fig:D} to be shown as contextual extensions rather than as inputs to the fitted $k$--$M_*$ relation. The redshift distributions also illustrate the distinct selection functions of the ELG, BGS, and LRG samples and the wavelength intervals over which the principal low-ionization transitions fall within the DESI spectral range. In particular, the Mg\,\textsc{ii} coverage requirement removes different fractions of the parent pair samples for the three target classes, as quantified in Section~\ref{sample}. These distributions are therefore useful for assessing the available statistical weight, redshift overlap, and spectral coverage of the stacks without entering directly into the interpretation of the radial trends.

\subsection{Stacked-profile diagnostics}\label{app:profiles}
\begin{figure*}[t]
    \centering
    \includegraphics[width=\textwidth]{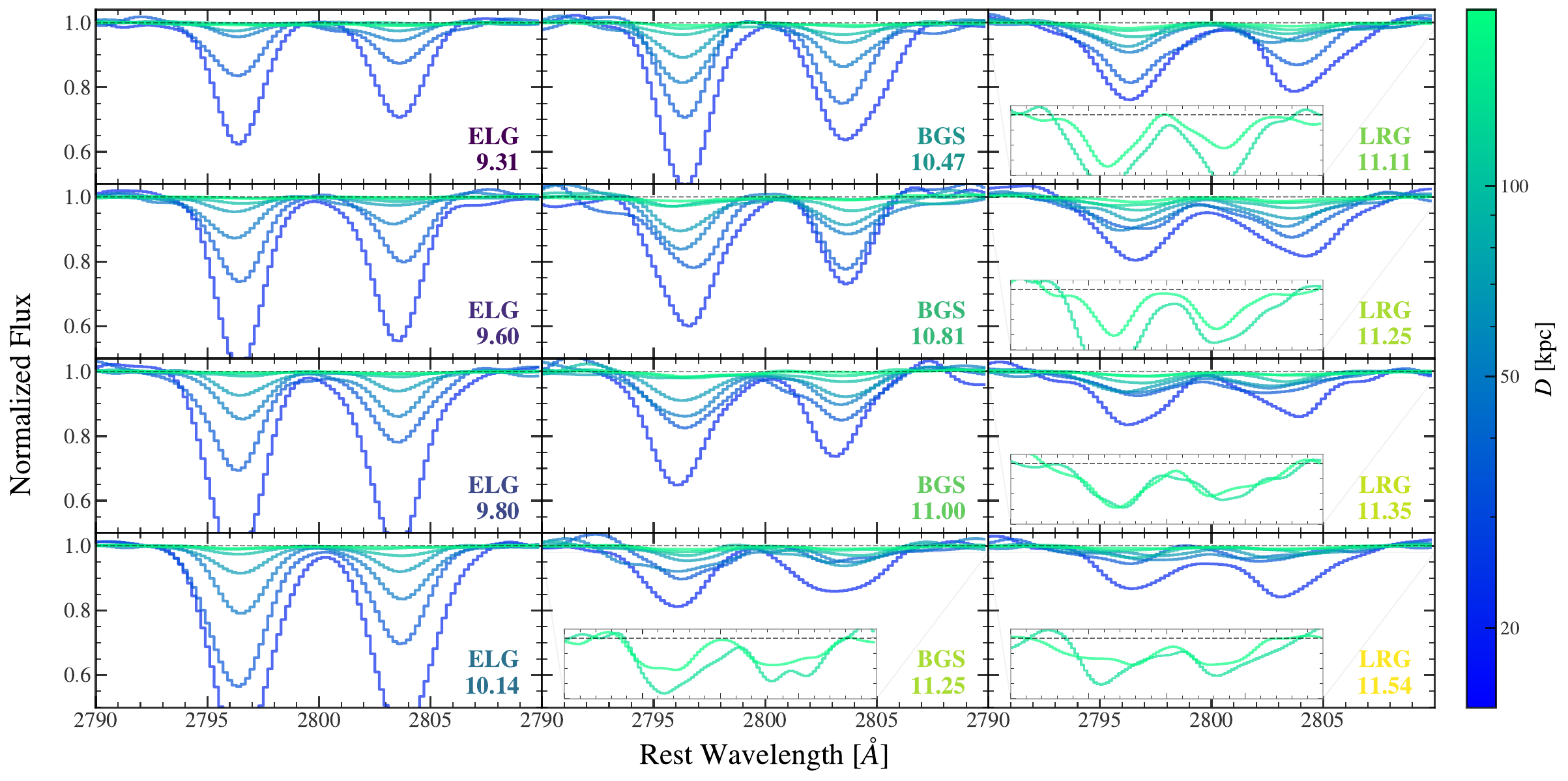}
    \caption{Mg\,\textsc{ii} absorption profiles across galaxy population and stellar mass. Columns show ELGs (left), BGSs (centre) and LRGs (right), and rows show stellar-mass bins; inset labels give $\log\langle M_*/\mathrm{M_{\odot}}\rangle$. Curves show Gaussian-smoothed normalized flux over rest-frame wavelengths of 2790--2810~\AA\ for seven projected-distance bins. Colours encode $D$ as indicated by the colour bar on the right. Insets in the bottom-row BGS and LRG panels magnify the two largest-$D$ profiles; the dashed line marks unit continuum.}
    \label{fig:specall}
\end{figure*}

Figure~\ref{fig:specall} presents the full set of stacked Mg\,\textsc{ii} profiles that underlie the radial equivalent-width measurements in Figure~\ref{fig:D}. The panels span the 12 stellar-mass populations used for the radial analysis, with the seven projected-distance bins shown directly in normalized-flux space. The progressive weakening of the Mg\,\textsc{ii} doublet with increasing projected distance is visible across the galaxy classes, while the detailed rate of decline varies with stellar mass, providing a direct spectral view of the profile-shape trend summarized by $k$. The two members of the doublet show consistent behaviour, and the enlarged insets for the largest-distance BGS and LRG stacks illustrate the regime in which the absorption approaches the continuum and the equivalent-width uncertainties become correspondingly more important. This figure is intended as a measurement diagnostic: it shows that the radial trends reported in the main text arise from coherent changes in the stacked absorption profiles rather than from a small number of isolated equivalent-width measurements.

\clearpage
\setcounter{figure}{0}
\setcounter{equation}{0}
\renewcommand{\thefigure}{\Alph{section}\arabic{figure}}
\renewcommand{\theHfigure}{appendix.\Alph{section}.\arabic{figure}}
\section{Robustness Tests}\label{app:robustness}
\subsection{Virial-radius normalization}\label{app:rvir}\label{rvirnorm}
\begin{figure*}[t]
    \centering
    \includegraphics[width=\textwidth]{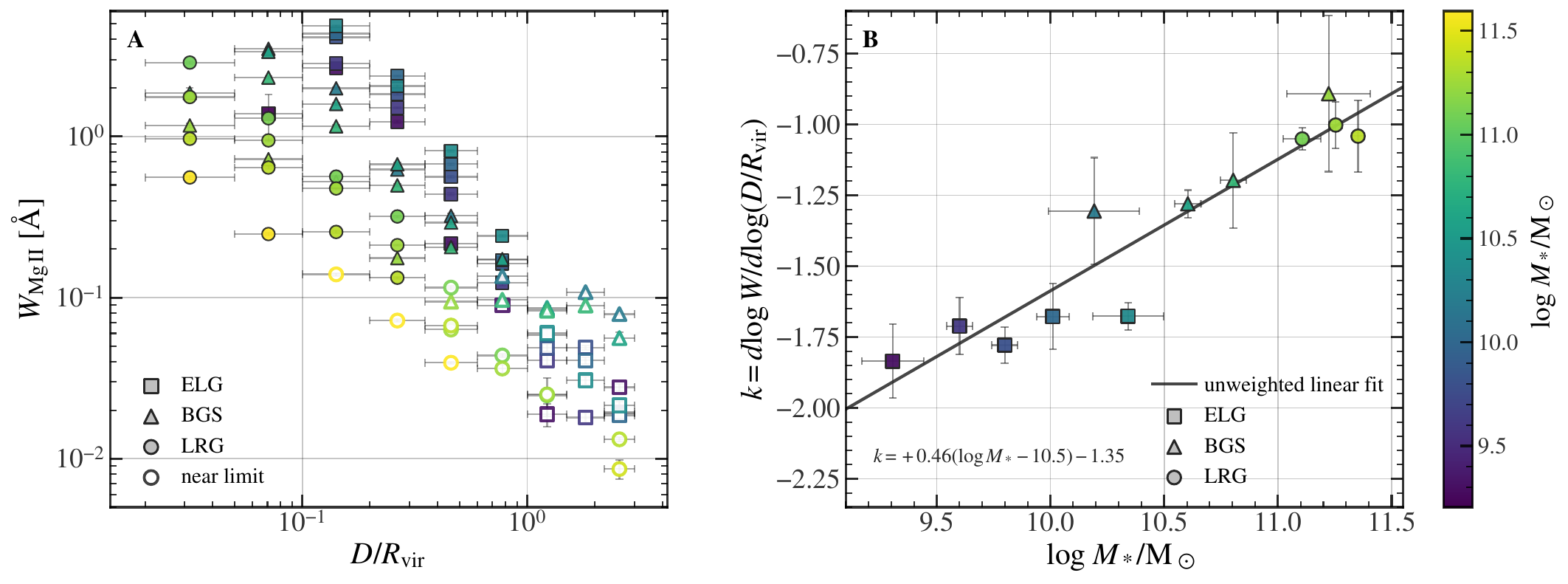}
    \caption{Mg\,\textsc{ii} equivalent width as a function of normalized projected distance. \textbf{(A)} The stacked $W_{\rm Mg\,II}$ measurements are shown against $D/R_{\rm vir}$. For this test, $R_{\rm vir}$ is evaluated for each galaxy and each galaxy--quasar pair is assigned to a radial bin using its own $D/R_{\rm vir}$ before stacking. This differs from the right panel of Figure~\ref{fig:D}, where the original fixed-$D$ stacks are only displayed relative to a representative bin-averaged virial radius. \textbf{(B)} The corresponding power-law slope $k=d\log W/d\log(D/R_{\rm vir})$ is shown as a function of stellar mass. Symbols indicate DESI target class and are colour-coded by stellar mass; open symbols in the left panel denote measurements close to the adopted detection or line-depth quality threshold. The persistence of the mass-dependent radial slope in the pair-level normalized stacks shows that the $k$--$M_*$ relation is not an artefact of using fixed physical-distance bins.}
    \label{fig:ewrvir}
\end{figure*}

To test explicitly whether the inferred radial trend is sensitive to the different halo sizes represented in a fixed-$D$ bin, we construct a second set of stacks in halo-normalized coordinates~\citep{2008ApJ...687..745C}. Recent DESI Y1 results suggest that radial slopes measured using fixed-$D$ and $D/R_{\rm vir}$ binning are broadly consistent, with only a minor halo-size effect~\citep{Kong2026}. Nevertheless, we test this explicitly for our sample. For each galaxy, $M_{\rm halo}$ is inferred from its stellar mass using the fiducial SHMR in Equation~\ref{eq:SHMR}, and $R_{\rm vir}$ is then evaluated at that galaxy's spectroscopic redshift through Equation~\ref{eq:Rv}. We compute $D/R_{\rm vir}$ for each galaxy--quasar pair before assigning the pair to a normalized radial bin. The values $\langle M_*\rangle$, $\langle z\rangle$, and $\langle R_{\rm vir}\rangle$ quoted for a resulting stack are therefore summary statistics of the contributing galaxies, not a single halo radius used to define the bin.

This procedure is methodologically distinct from the right panel of Figure~\ref{fig:D}. There, the spectra are first stacked in the same fixed physical-distance bins used in the left panel, and the resulting radial coordinate is subsequently displayed relative to the representative $\langle R_{\rm vir}\rangle$ of the corresponding stellar-mass bin. In Figure~\ref{fig:ewrvir}, by contrast, the normalization is performed pair by pair before radial binning and spectral stacking. The persistence of the mass-dependent slope $k$ in the pair-level $D/R_{\rm vir}$ stacks shows that the observed $k$--$M_*$ relation is not an artefact of using fixed physical-distance bins.

\subsection{Ion and redshift robustness of the radial slope}\label{app:krobust}\label{robustness}
\begin{figure*}[t]
    \centering
    \includegraphics[width=\textwidth]{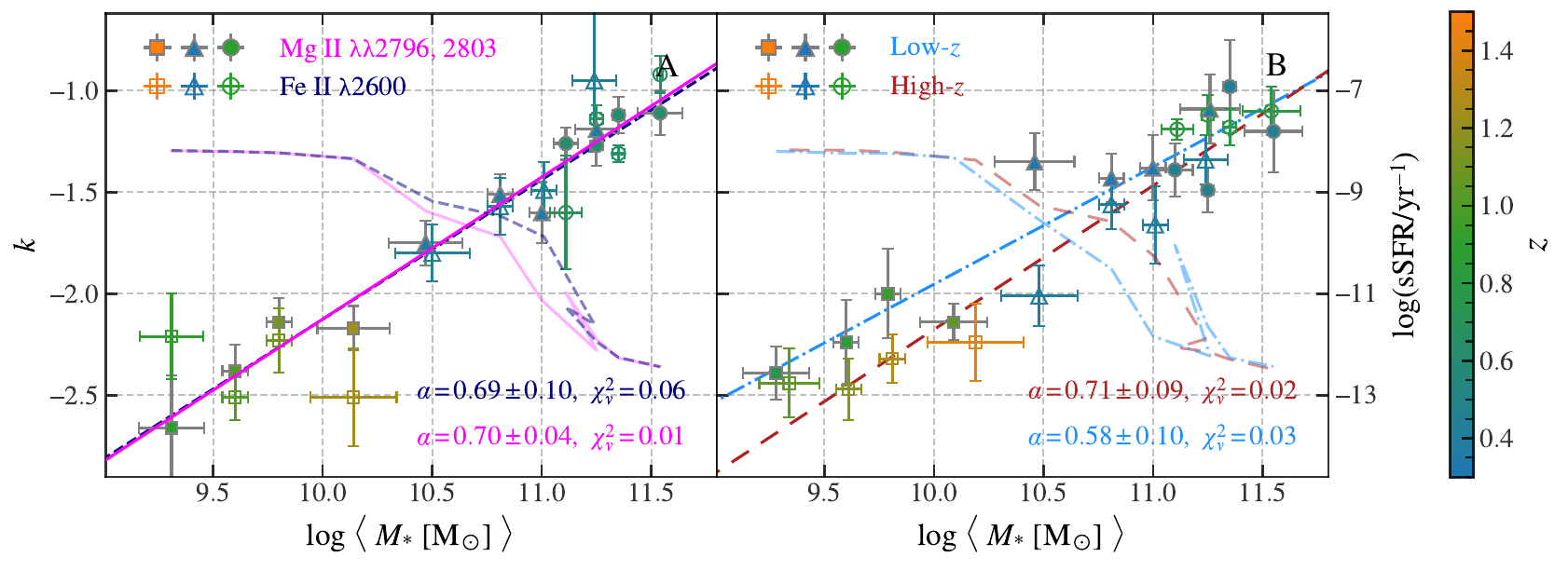}
    \caption{Robustness of the radial-slope--stellar-mass relation. \textbf{(A)} slopes $k$ of the equivalent-width--distance relations measured from Mg\,\textsc{ii}\,$\lambda\lambda2796,2803$ (filled symbols) and Fe\,\textsc{ii}\,$\lambda2600$ (open symbols) stacks as a function of stellar mass. \textbf{(B)} each Mg\,\textsc{ii} stellar-mass bin is divided at its median redshift; filled and open symbols denote the lower- and higher-redshift halves, respectively. The quantity $a_k$ shown in the panel denotes the descriptive slope of a plotted $k$--$\log M_*$ fit, $a_k=d k/d\log M_*$. The coloured fit lines are shown only as visual summaries of the plotted subsamples; because the DESI target classes occupy different stellar-mass and redshift ranges, our robustness inference is based primarily on the within-bin comparison of the low- and high-redshift measurements rather than on interpreting a single global fit across target classes. The broken curves are retained only to guide the eye through the median $\log({\rm sSFR})$ values associated with the plotted subsamples. Apparent bends can occur where the sequence passes between differently selected DESI target classes; the curves are not fitted and are not interpreted as a continuous physical $\mathrm{sSFR}$--mass relation. Mg\,\textsc{ii} and Fe\,\textsc{ii} exhibit similar $k$--$M_*$ behaviour, and the redshift-split measurements show that the radial-slope trend remains visible when redshift differences are tracked.}\label{fig:k-M}
\end{figure*}

To test whether the mass dependence of the radial slope is specific to Mg\,\textsc{ii} or driven by redshift mixing, we extend the analysis to Fe\,\textsc{ii}\,$\lambda2600$ and split each Mg\,\textsc{ii} stellar-mass bin at its median redshift (Figure~\ref{fig:k-M}). Fe\,\textsc{ii} exhibits a similar $k$--$M_*$ trend to Mg\,\textsc{ii}, indicating that the mass dependence is not peculiar to a single ionic tracer but reflects the radial structure of the low-ionization cool CGM. Within individual stellar-mass bins, the low- and high-redshift subsamples yield consistent $k$ values within the uncertainties, including the ELG sample with the broadest redshift range. Together, these tests show that the radial-slope trend is primarily governed by stellar mass rather than redshift, and is consistently traced by both Mg\,\textsc{ii} and Fe\,\textsc{ii}.

\subsection{Redshift and sSFR sensitivity of the absorption}\label{app:redshift-ssfr}
\begin{figure*}[t]
    \centering
    \includegraphics[width=\textwidth]{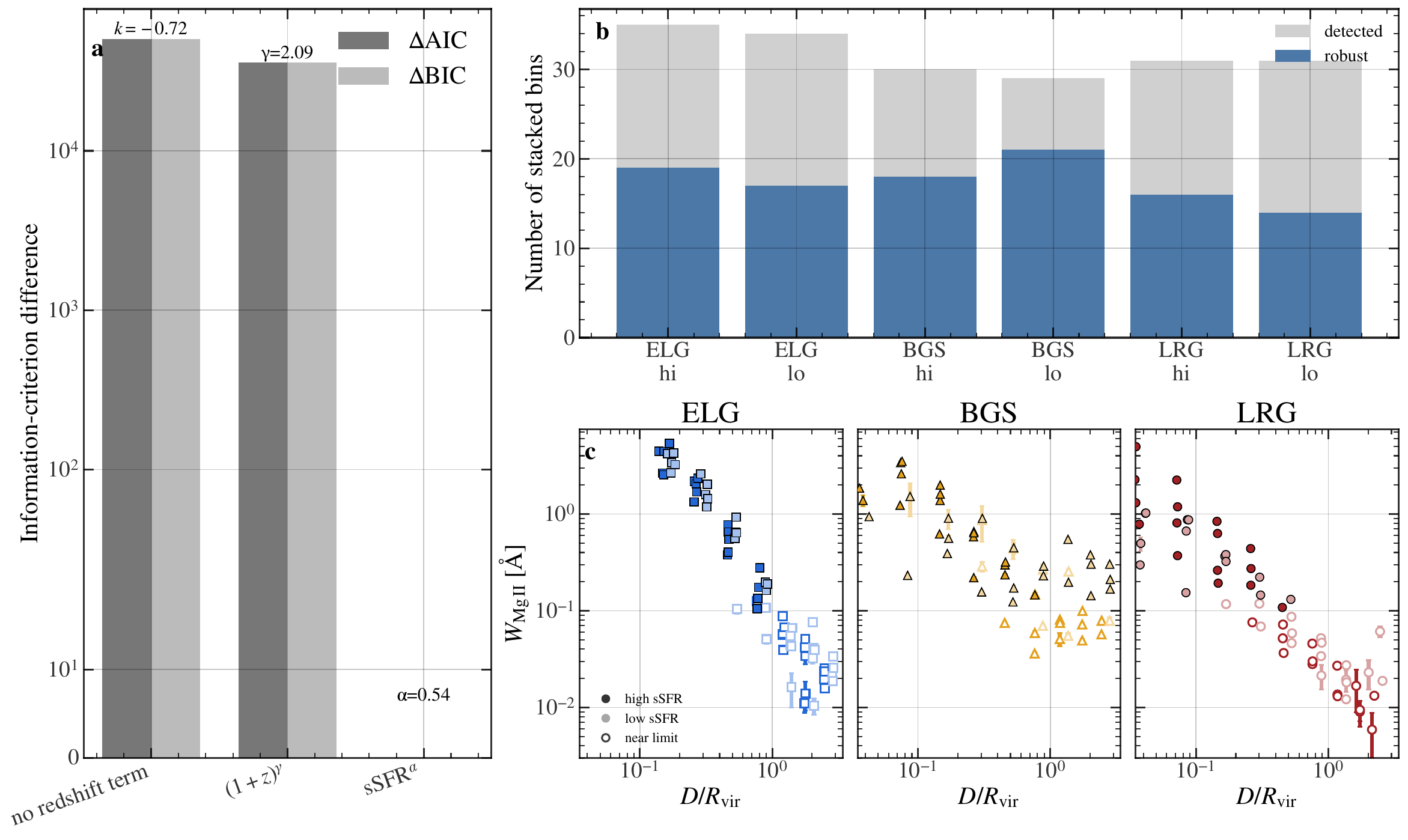}
    \caption{Robustness of the radial-slope--stellar-mass relation.
\textbf{(A)} Power-law slopes $k$ of the equivalent-width--distance relations measured from Mg\,\textsc{ii}\,$\lambda\lambda2796,2803$ (filled symbols) and Fe\,\textsc{ii}\,$\lambda2600$ (open symbols) stacks as a function of stellar mass.
\textbf{(B)} The Mg\,\textsc{ii} measurements split at the median redshift within each stellar-mass bin; filled and open symbols denote the lower- and higher-redshift subsamples, respectively.
The quantity $a_k=d k/d\log M_*$ gives the descriptive slope of the plotted $k$--$\log M_*$ relation. Coloured lines are shown as visual summaries, while the broken curves trace the median $\log(\mathrm{sSFR})$ values for guidance only.
The similar Mg\,\textsc{ii} and Fe\,\textsc{ii} trends, together with the consistency of the redshift-split measurements, show that the radial-slope variation is primarily associated with stellar mass rather than redshift.}
    \label{fig:modelselection}
\end{figure*}

\begin{figure*}[t]
    \centering
    \includegraphics[width=\textwidth]{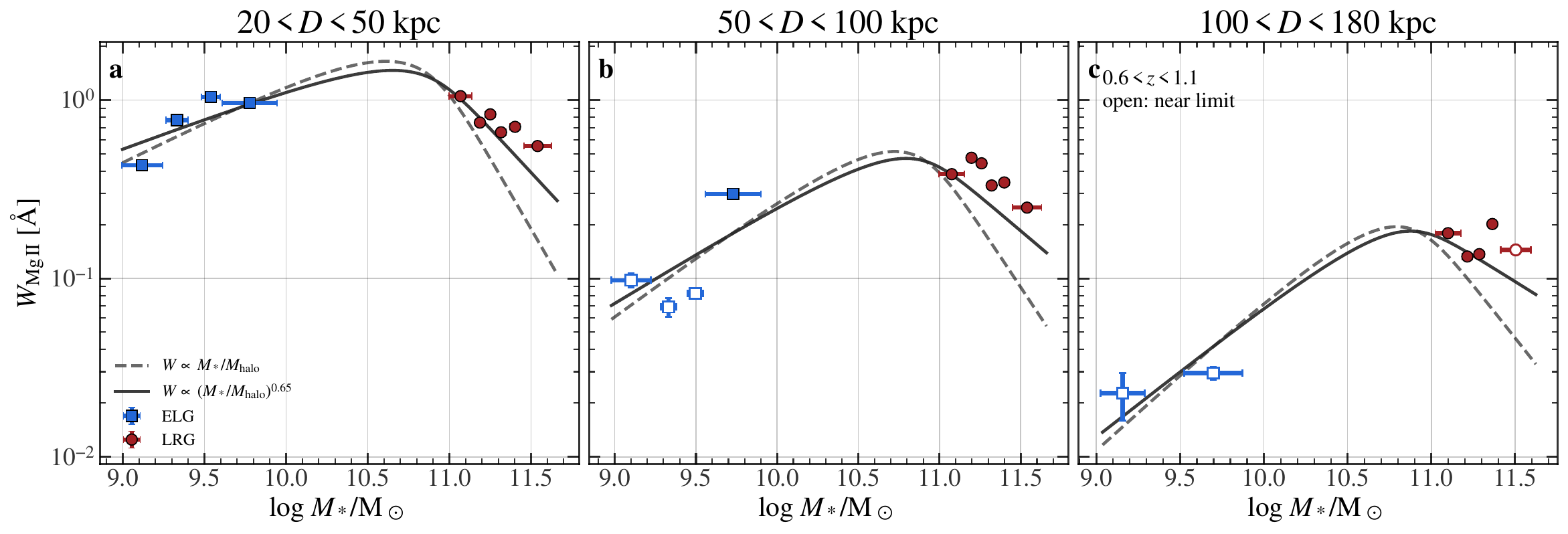}
    \caption{Redshift-controlled ELG/LRG Mg\,\textsc{ii} stacks. The $W_{\rm MgII}$--$M_*$ measurements are repeated using only $0.6<z<1.1$, where ELGs and LRGs substantially overlap in redshift. Panels \textbf{(A)}, \textbf{(B)}, and \textbf{(C)} correspond to $20<D<50$, $50<D<100$, and $100<D<180$~kpc, respectively. Filled points satisfy the conservative signal-to-noise and doublet-depth cuts; open points lie near the adopted detection or quality limit. Dashed and solid curves reproduce the $W_{\rm MgII}\propto M_*/M_{\rm halo}$ and $W_{\rm MgII}\propto(M_*/M_{\rm halo})^{0.65}$ empirical relations from Figure~\ref{fig:normalization}, carried over without refitting to the restricted-redshift sample. BGS galaxies are omitted because their useful Mg\,\textsc{ii} redshift coverage is largely disjoint from this interval.}
    \label{fig:zoverlap}
\end{figure*}

The DESI target classes span different redshift and star-formation regimes, so we perform two additional tests of the Mg\,\textsc{ii} absorption normalization and radial signal. First, we fit three empirical models to the same quality-selected $D/R_{\rm vir}$ stacks. Defining $x\equiv D/R_{\rm vir}$, the models are
\begin{equation}
\begin{array}{rcl}
W_{\rm MgII}^{\rm base} & = & A\,x^k,\\
W_{\rm MgII}^{z} & = & A\,x^k\left[(1+z)/(1+z_0)\right]^{\gamma},\\
W_{\rm MgII}^{\rm sSFR} & = & A\,x^k\left({\rm sSFR}/{\rm sSFR}_0\right)^{\alpha},
\end{array}
\end{equation}
where $z_0=0.7$ and ${\rm sSFR}_0=10^{-9.5}\,{\rm yr}^{-1}$ are fixed pivots. The baseline model has two free parameters, $(A,k)$, and the other models add either $\gamma$ or $\alpha$. The redshift-dependent model improves upon the baseline fit, with $\gamma=2.09$, while the sSFR-dependent model has the lowest AIC and BIC among the tested forms, with $\alpha=0.54$ and a residual scatter of 0.25~dex (Figure~\ref{fig:modelselection}). When each target class and stellar-mass bin is divided at its median sSFR, both halves retain measurable radial declines. The radial signal therefore persists after subdivision by current star-formation state, while the model comparison shows that sSFR carries additional information about the absorption normalization.

Second, we repeat the $W_{\rm MgII}$--$M_*$ measurement using only $0.6<z<1.1$, where the ELG and LRG samples overlap substantially in redshift (Figure~\ref{fig:zoverlap}). BGS galaxies are omitted because their useful Mg\,\textsc{ii} redshift coverage is largely disjoint from this interval. In all three projected-distance ranges, the restricted-redshift stacks retain the low-mass rise, broad intermediate-mass maximum, and high-mass decline seen in the full sample. The two SHMR-based empirical relations from Figure~\ref{fig:normalization} are overplotted without refitting. The persistence of the turnover in this common-redshift interval shows that the observed mass dependence is not produced solely by the different redshift distributions of the DESI target classes.

\clearpage
\setcounter{figure}{0}
\setcounter{equation}{0}
\section{Extended Comparisons}\label{app:extended}
\subsection{Comparison with previous work}\label{app:previous}\label{previous}
\begin{figure}[t]
    \centering
    \includegraphics[width=\columnwidth]{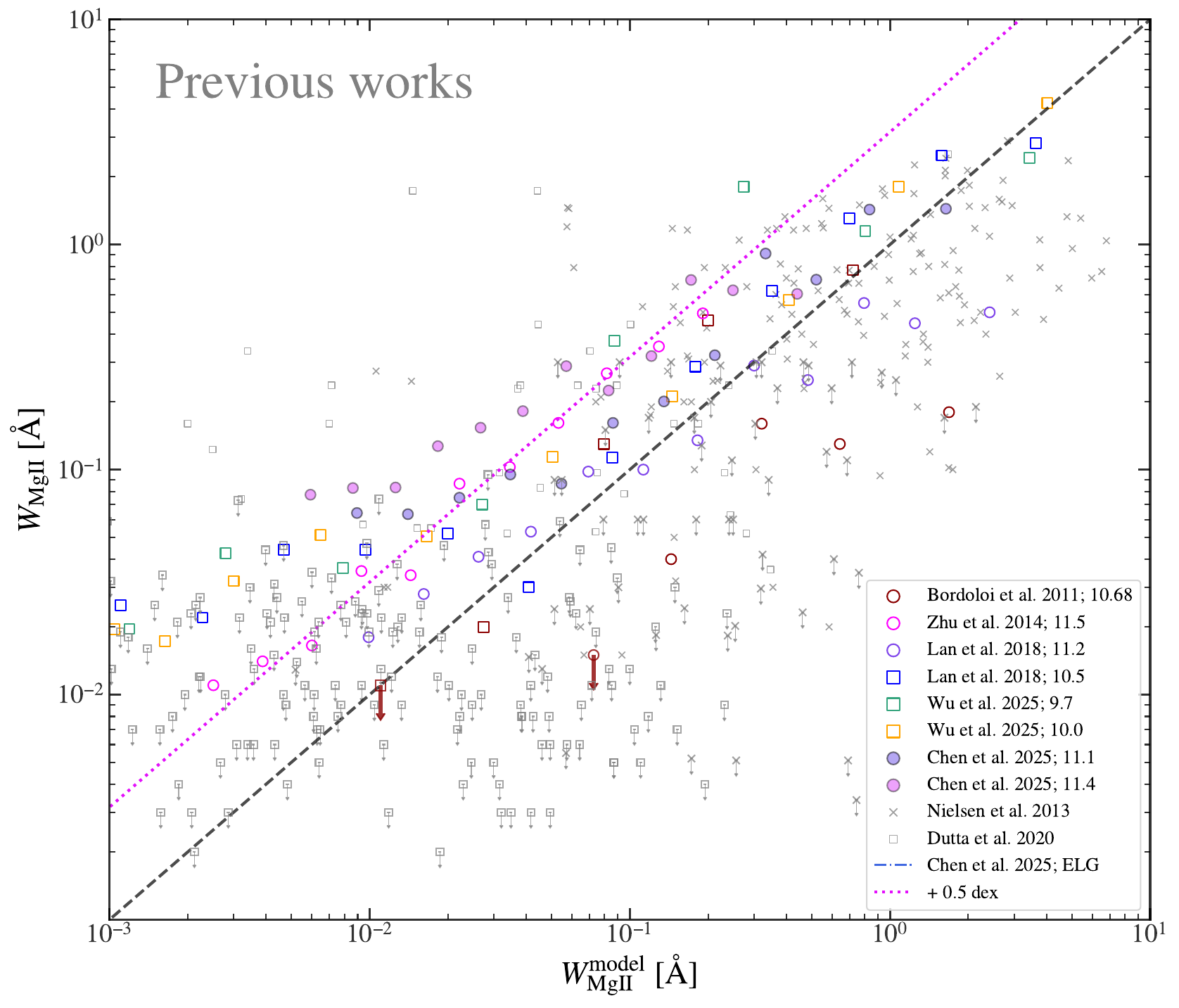}
    \caption{Comparison of observed and modelled Mg\,\textsc{ii} equivalent widths. Literature measurements of $W_{\rm MgII}$ are shown against the prediction $W_{\rm model}$ from Equation~\ref{eq:QR}. Symbols denote \citet{bordoloi11} (dark-red open squares and circles), \citet{zhu14} (magenta open circles), \citet{lan18} (blue open squares and circles), \citet{wu2024} (green and orange open squares), \citet{chen25} (filled circles and blue dotted curves), \citet{nielsen13,nielsen13b} (grey crosses; individual absorbers) and \citet{dutta20} (grey open squares; individual absorbers). The dotted magenta line marks a 0.5-dex excess above the model prediction. Squares and circles denote ELG and LRG measurements, respectively. Blue curves show the best-fitting relations within 100~kpc for ELGs with stellar masses of $10^{8.8}$, $10^{9.8}$ and $10^{10.5}\,\mathrm{M_\odot}$ from \citet{chen25}. Numeric labels give $\log(M_*/\mathrm{M_\odot})$ for each sample. Stellar masses from \citet{nielsen13,nielsen13b} are converted from absolute $K$-band magnitude using $\log(M_*/\mathrm{M_\odot})=-0.4(M_K-3.28)+\log0.6$~\citep{McGaugh14}.}
    \label{fig:compare}
\end{figure}

In Figure~\ref{fig:compare}, we compare the empirical model in Equation~\ref{eq:QR} with previous Mg\,\textsc{ii} measurements spanning a broad range of stellar mass, projected distance and measurement strategy. The model provides a useful first-order description of several literature measurements~\citep{zhu14,lan18,chen25,wu2024} and lies within much of the scatter of individual-absorber measurements~\citep{nielsen13,nielsen13b,dutta20}. This agreement should not be interpreted as evidence that the parametrization captures a unique underlying physical mechanism; rather, it shows that a common mass--radius scaling can summarize a heterogeneous set of Mg\,\textsc{ii} measurements. Systematic deviations become more evident for LRGs, especially at the high-mass end, where measurements~\citep{zhu14,chen25} lie above the fiducial prediction. Intermediate-mass LRG measurements~\citep{lan18,chen25} are closer to the model, whereas the lower-mass sample of \citet{bordoloi11} is somewhat weaker. The comparison therefore supports using the SHMR-based scaling as an empirical baseline, while indicating that additional population-dependent terms are needed to describe the massive, quenched regime.

\clearpage
\bibliographystyle{aasjournalv7.1}
\bibliography{maix}

@ARTICLE{Muratov-15,
       author = {{Muratov}, Alexander L. and {Kere{\v{s}}}, Du{\v{s}}an and {Faucher-Gigu{\`e}re}, Claude-Andr{\'e} and {Hopkins}, Philip F. and {Quataert}, Eliot and {Murray}, Norman},
        title = "{Gusty, gaseous flows of FIRE: galactic winds in cosmological simulations with explicit stellar feedback}",
      journal = {\mnras},
         year = 2015,
        month = dec,
       volume = {454},
       number = {3},
        pages = {2691-2713},
          doi = {10.1093/mnras/stv2126},
archivePrefix = {arXiv},
       eprint = {1501.03155},
 primaryClass = {astro-ph.GA},
       adsurl = {https://ui.adsabs.harvard.edu/abs/2015MNRAS.454.2691M}
}

@ARTICLE{Lyu-26,
       author = {{Lyu}, Cheqiu and {Wang}, Enci and {Chen}, Zeyu and {Ma}, Chengyu and {Chen}, Yangyao and {Yu}, Haoran and {Kong}, Xu},
        title = "{Observational Evidence for Anisotropic Metal Excess around Galaxies}",
      journal = {arXiv e-prints},
         year = 2026,
        month = jul,
          eid = {arXiv:2607.24929},
        pages = {arXiv:2607.24929},
          doi = {10.48550/arXiv.2607.24929},
archivePrefix = {arXiv},
       eprint = {2607.24929},
 primaryClass = {astro-ph.GA},
       adsurl = {https://ui.adsabs.harvard.edu/abs/2026arXiv260724929L}
}

@ARTICLE{Ma-26,
       author = {{Ma}, Chengyu and {Chen}, Yangyao and {Wang}, Enci and {Mo}, Houjun and {Wang}, Huiyuan and {Wang}, Tao and {Yuan}, Qirong and {Du}, Min and {Li}, Zhaozhou and {Wang}, Kai and {Lyu}, Cheqiu and {Yu}, Haoran and {Chen}, Zeyu and {Chen}, Xiaoxuan},
        title = "{Fountain pattern of baryon cycle revealed in galaxy ecosystems}",
      journal = {arXiv e-prints},
         year = 2026,
        month = sep,
          eid = {arXiv:2609.21374},
        pages = {arXiv:2609.21374},
          doi = {10.48550/arXiv.2609.21374},
archivePrefix = {arXiv},
       eprint = {2609.21374},
 primaryClass = {astro-ph.GA},
       adsurl = {https://ui.adsabs.harvard.edu/abs/2026arXiv260921374M}
}

@ARTICLE{Kong2026,
       author = {{Kong}, Weiwen and {Chen}, Zeyu and {Wang}, Enci and {Yu}, Haoran and {Wang}, Kai and {Shi}, Dongdong and {Lyu}, Cheqiu and {Zhang}, Yuxuan and {Zhang}, Haoyi and {Guan}, Haowen},
        title = "{Cooler Phases of the Circumgalactic Medium Are More Centrally Concentrated: Constraints from Multiphase Absorption Lines}",
      journal = {\apj},
         year = 2026,
        month = jul,
       volume = {1006},
       number = {1},
          eid = {49},
        pages = {49},
          doi = {10.3847/1538-4357/ae80b0},
archivePrefix = {arXiv},
       eprint = {2606.04793},
 primaryClass = {astro-ph.GA},
       adsurl = {https://ui.adsabs.harvard.edu/abs/2026ApJ..1006...49K}
}

@ARTICLE{2016A&A...594A..13P,
       author = {{Planck Collaboration} and {Ade}, P.~A.~R. and {Aghanim}, N. and {Arnaud}, M. and {Ashdown}, M. and {Aumont}, J. and {Baccigalupi}, C. and {Banday}, A.~J. and {Barreiro}, R.~B. and {Bartlett}, J.~G. and {Bartolo}, N. and {Battaner}, E. and {Battye}, R. and {Benabed}, K. and {Beno{\^\i}t}, A. and {Benoit-L{\'e}vy}, A. and {Bernard}, J. -P. and {Bersanelli}, M. and {Bielewicz}, P. and {Bock}, J.~J. and {Bonaldi}, A. and {Bonavera}, L. and {Bond}, J.~R. and {Borrill}, J. and {Bouchet}, F.~R. and {Boulanger}, F. and {Bucher}, M. and {Burigana}, C. and {Butler}, R.~C. and {Calabrese}, E. and {Cardoso}, J. -F. and {Catalano}, A. and {Challinor}, A. and {Chamballu}, A. and {Chary}, R. -R. and {Chiang}, H.~C. and {Chluba}, J. and {Christensen}, P.~R. and {Church}, S. and {Clements}, D.~L. and {Colombi}, S. and {Colombo}, L.~P.~L. and {Combet}, C. and {Coulais}, A. and {Crill}, B.~P. and {Curto}, A. and {Cuttaia}, F. and {Danese}, L. and {Davies}, R.~D. and {Davis}, R.~J. and {de Bernardis}, P. and {de Rosa}, A. and {de Zotti}, G. and {Delabrouille}, J. and {D{\'e}sert}, F. -X. and {Di Valentino}, E. and {Dickinson}, C. and {Diego}, J.~M. and {Dolag}, K. and {Dole}, H. and {Donzelli}, S. and {Dor{\'e}}, O. and {Douspis}, M. and {Ducout}, A. and {Dunkley}, J. and {Dupac}, X. and {Efstathiou}, G. and {Elsner}, F. and {En{\ss}lin}, T.~A. and {Eriksen}, H.~K. and {Farhang}, M. and {Fergusson}, J. and {Finelli}, F. and {Forni}, O. and {Frailis}, M. and {Fraisse}, A.~A. and {Franceschi}, E. and {Frejsel}, A. and {Galeotta}, S. and {Galli}, S. and {Ganga}, K. and {Gauthier}, C. and {Gerbino}, M. and {Ghosh}, T. and {Giard}, M. and {Giraud-H{\'e}raud}, Y. and {Giusarma}, E. and {Gjerl{\o}w}, E. and {Gonz{\'a}lez-Nuevo}, J. and {G{\'o}rski}, K.~M. and {Gratton}, S. and {Gregorio}, A. and {Gruppuso}, A. and {Gudmundsson}, J.~E. and {Hamann}, J. and {Hansen}, F.~K. and {Hanson}, D. and {Harrison}, D.~L. and {Helou}, G. and {Henrot-Versill{\'e}}, S. and {Hern{\'a}ndez-Monteagudo}, C. and {Herranz}, D. and {Hildebrandt}, S.~R. and {Hivon}, E. and {Hobson}, M. and {Holmes}, W.~A. and {Hornstrup}, A. and {Hovest}, W. and {Huang}, Z. and {Huffenberger}, K.~M. and {Hurier}, G. and {Jaffe}, A.~H. and {Jaffe}, T.~R. and {Jones}, W.~C. and {Juvela}, M. and {Keih{\"a}nen}, E. and {Keskitalo}, R. and {Kisner}, T.~S. and {Kneissl}, R. and {Knoche}, J. and {Knox}, L. and {Kunz}, M. and {Kurki-Suonio}, H. and {Lagache}, G. and {L{\"a}hteenm{\"a}ki}, A. and {Lamarre}, J. -M. and {Lasenby}, A. and {Lattanzi}, M. and {Lawrence}, C.~R. and {Leahy}, J.~P. and {Leonardi}, R. and {Lesgourgues}, J. and {Levrier}, F. and {Lewis}, A. and {Liguori}, M. and {Lilje}, P.~B. and {Linden-V{\o}rnle}, M. and {L{\'o}pez-Caniego}, M. and {Lubin}, P.~M. and {Mac{\'\i}as-P{\'e}rez}, J.~F. and {Maggio}, G. and {Maino}, D. and {Mandolesi}, N. and {Mangilli}, A. and {Marchini}, A. and {Maris}, M. and {Martin}, P.~G. and {Martinelli}, M. and {Mart{\'\i}nez-Gonz{\'a}lez}, E. and {Masi}, S. and {Matarrese}, S. and {McGehee}, P. and {Meinhold}, P.~R. and {Melchiorri}, A. and {Melin}, J. -B. and {Mendes}, L. and {Mennella}, A. and {Migliaccio}, M. and {Millea}, M. and {Mitra}, S. and {Miville-Desch{\^e}nes}, M. -A. and {Moneti}, A. and {Montier}, L. and {Morgante}, G. and {Mortlock}, D. and {Moss}, A. and {Munshi}, D. and {Murphy}, J.~A. and {Naselsky}, P. and {Nati}, F. and {Natoli}, P. and {Netterfield}, C.~B. and {N{\o}rgaard-Nielsen}, H.~U. and {Noviello}, F. and {Novikov}, D. and {Novikov}, I. and {Oxborrow}, C.~A. and {Paci}, F. and {Pagano}, L. and {Pajot}, F. and {Paladini}, R. and {Paoletti}, D. and {Partridge}, B. and {Pasian}, F. and {Patanchon}, G. and {Pearson}, T.~J. and {Perdereau}, O. and {Perotto}, L. and {Perrotta}, F. and {Pettorino}, V. and {Piacentini}, F. and {Piat}, M. and {Pierpaoli}, E. and {Pietrobon}, D. and {Plaszczynski}, S. and {Pointecouteau}, E. and {Polenta}, G. and {Popa}, L. and {Pratt}, G.~W. and {Pr{\'e}zeau}, G. and {Prunet}, S. and {Puget}, J. -L. and {Rachen}, J.~P. and {Reach}, W.~T. and {Rebolo}, R. and {Reinecke}, M. and {Remazeilles}, M. and {Renault}, C. and {Renzi}, A. and {Ristorcelli}, I. and {Rocha}, G. and {Rosset}, C. and {Rossetti}, M. and {Roudier}, G. and {Rouill{\'e} d'Orfeuil}, B. and {Rowan-Robinson}, M. and {Rubi{\~n}o-Mart{\'\i}n}, J.~A. and {Rusholme}, B. and {Said}, N. and {Salvatelli}, V. and {Salvati}, L. and {Sandri}, M. and {Santos}, D. and {Savelainen}, M. and {Savini}, G. and {Scott}, D. and {Seiffert}, M.~D. and {Serra}, P. and {Shellard}, E.~P.~S. and {Spencer}, L.~D. and {Spinelli}, M. and {Stolyarov}, V. and {Stompor}, R. and {Sudiwala}, R. and {Sunyaev}, R. and {Sutton}, D. and {Suur-Uski}, A. -S. and {Sygnet}, J. -F. and {Tauber}, J.~A. and {Terenzi}, L. and {Toffolatti}, L. and {Tomasi}, M. and {Tristram}, M. and {Trombetti}, T. and {Tucci}, M. and {Tuovinen}, J. and {T{\"u}rler}, M. and {Umana}, G. and {Valenziano}, L. and {Valiviita}, J. and {Van Tent}, F. and {Vielva}, P. and {Villa}, F. and {Wade}, L.~A. and {Wandelt}, B.~D. and {Wehus}, I.~K. and {White}, M. and {White}, S.~D.~M. and {Wilkinson}, A. and {Yvon}, D. and {Zacchei}, A. and {Zonca}, A.},
        title = "{Planck 2015 results. XIII. Cosmological parameters}",
      journal = {\aap},
         year = 2016,
        month = sep,
       volume = {594},
          eid = {A13},
        pages = {A13},
          doi = {10.1051/0004-6361/201525830},
archivePrefix = {arXiv},
       eprint = {1502.01589},
 primaryClass = {astro-ph.CO},
       adsurl = {https://ui.adsabs.harvard.edu/abs/2016A&A...594A..13P}
}

@article{zhu13,
  title={The JHU-SDSS metal absorption line catalog: redshift evolution and properties of Mg II absorbers},
  author={Zhu, Guangtun and M{\'e}nard, Brice},
  journal={\apj},
  volume={770},
  number={2},
  pages={130},
  year={2013},
  publisher={IOP Publishing},
  doi={10.1088/0004-637X/770/2/130},
  archivePrefix={arXiv},
  eprint={1211.6215},
  primaryClass={astro-ph.CO},
  adsurl={https://ui.adsabs.harvard.edu/abs/2013ApJ...770..130Z}
}

@article{lan18,
  title={The circumgalactic medium of eBOSS emission line galaxies: signatures of galactic outflows in gas distribution and kinematics},
  author={Lan, Ting-Wen and Mo, Houjun},
  journal={\apj},
  volume={866},
  number={1},
  pages={36},
  year={2018},
  publisher={IOP Publishing},
  doi={10.3847/1538-4357/aadc08},
  archivePrefix={arXiv},
  eprint={1806.05786},
  primaryClass={astro-ph.GA},
  adsurl={https://ui.adsabs.harvard.edu/abs/2018ApJ...866...36L}
}

@article{dutta20,
  title={MUSE Analysis of Gas around Galaxies (MAGG)--II: metal-enriched halo gas around z~ 1 galaxies},
  author={Dutta, Rajeshwari and Fumagalli, Michele and Fossati, Matteo and Lofthouse, Emma K and Prochaska, J Xavier and Arrigoni Battaia, Fabrizio and Bielby, Richard M and Cantalupo, Sebastiano and Cooke, Ryan J and Murphy, Michael T and others},
  journal={\mnras},
  volume={499},
  number={4},
  pages={5022--5046},
  year={2020},
  publisher={Oxford University Press},
  doi={10.1093/mnras/staa3147},
  archivePrefix={arXiv},
  eprint={2009.14219},
  primaryClass={astro-ph.GA},
  adsurl={https://ui.adsabs.harvard.edu/abs/2020MNRAS.499.5022D}
}

@article{nielsen13,
  title={MAGiiCAT I. THE Mg ii ABSORBER--GALAXY CATALOG},
  author={Nielsen, Nikole M and Churchill, Christopher W and Kacprzak, Glenn G and Murphy, Michael T},
  journal={\apj},
  volume={776},
  number={2},
  pages={114},
  year={2013},
  publisher={IOP Publishing},
  doi={10.1088/0004-637X/776/2/114},
  archivePrefix={arXiv},
  eprint={1304.6716},
  primaryClass={astro-ph.CO},
  adsurl={https://ui.adsabs.harvard.edu/abs/2013ApJ...776..114N}
}

@article{nielsen13b,
  title={MAGIICAT II. General characteristics of the Mg II absorbing circumgalactic medium},
  author={Nielsen, Nikole M and Churchill, Christopher W and Kacprzak, Glenn G},
  journal={\apj},
  volume={776},
  number={2},
  pages={115},
  year={2013},
  publisher={IOP Publishing},
  doi={10.1088/0004-637X/776/2/115},
  archivePrefix={arXiv},
  eprint={1211.1380},
  primaryClass={astro-ph.CO},
  adsurl={https://ui.adsabs.harvard.edu/abs/2013ApJ...776..115N}
}

@article{bordoloi11,
  title={The radial and azimuthal profiles of Mg II absorption around 0.5< z< 0.9 zCOSMOS galaxies of different colors, masses, and environments},
  author={Bordoloi, Rongmon and Lilly, Simon J and Knobel, Christian and Bolzonella, Micoll and Kampczyk, Pawel and Carollo, C Marcella and Iovino, Angela and Zucca, Elena and Contini, Thierry and Kneib, J-P and others},
  journal={\apj},
  volume={743},
  number={1},
  pages={10},
  year={2011},
  publisher={IOP Publishing},
  doi={10.1088/0004-637X/743/1/10},
  archivePrefix={arXiv},
  eprint={1106.0616},
  primaryClass={astro-ph.CO},
  adsurl={https://ui.adsabs.harvard.edu/abs/2011ApJ...743...10B}
}

@article{myers23,
  title={The Target-selection Pipeline for the Dark Energy Spectroscopic Instrument},
  author={Myers, Adam D and Moustakas, John and Bailey, Stephen and Weaver, Benjamin A and Cooper, Andrew P and Forero-Romero, Jaime E and Abolfathi, Bela and Alexander, David M and Brooks, David and Chaussidon, Edmond and others},
  journal={\aj},
  volume={165},
  number={2},
  pages={50},
  year={2023},
  publisher={IOP Publishing},
  doi={10.3847/1538-3881/aca5f9},
  archivePrefix={arXiv},
  eprint={2208.08518},
  primaryClass={astro-ph.IM},
  adsurl={https://ui.adsabs.harvard.edu/abs/2023AJ....165...50M}
}

@misc{guo18,
  title={PyQSOFit: Python code to fit the spectrum of quasars},
  author={Guo, Hengxiao and Shen, Yue and Wang, Shu},
  howpublished={Astrophysics Source Code Library, record ascl:1809.008},
  year={2018},
  month={sep},
  eid={ascl:1809.008},
  eprint={1809.008},
  adsurl={https://ui.adsabs.harvard.edu/abs/2018ascl.soft09008G}
}

@article{tumlinson17,
  title={The circumgalactic medium},
  author={Tumlinson, Jason and Peeples, Molly S and Werk, Jessica K},
  journal={\araa},
  volume={55},
  pages={389--432},
  year={2017},
  publisher={Annual Reviews},
  doi={10.1146/annurev-astro-091916-055240},
  archivePrefix={arXiv},
  eprint={1709.09180},
  primaryClass={astro-ph.GA},
  adsurl={https://ui.adsabs.harvard.edu/abs/2017ARA&A..55..389T}
}

@article{zhu13ca,
  title={Calcium H \& K induced by galaxy halos},
  author={Zhu, Guangtun and M{\'e}nard, Brice},
  journal={\apj},
  volume={773},
  number={1},
  pages={16},
  year={2013},
  publisher={IOP Publishing},
  doi={10.1088/0004-637X/773/1/16},
  archivePrefix={arXiv},
  eprint={1304.0451},
  primaryClass={astro-ph.CO},
  adsurl={https://ui.adsabs.harvard.edu/abs/2013ApJ...773...16Z}
}

@article{bordoloi14,
  title={The dependence of galactic outflows on the properties and orientation of zCOSMOS galaxies at z~ 1},
  author={Bordoloi, R and Lilly, SJ and Hardmeier, E and Contini, T and Kneib, J-P and Le F{\`e}vre, Olivier and Mainieri, V and Renzini, A and Scodeggio, M and Zamorani, G and others},
  journal={\apj},
  volume={794},
  number={2},
  pages={130},
  year={2014},
  publisher={IOP Publishing},
  doi={10.1088/0004-637X/794/2/130},
  archivePrefix={arXiv},
  eprint={1307.6553},
  primaryClass={astro-ph.CO},
  adsurl={https://ui.adsabs.harvard.edu/abs/2014ApJ...794..130B}
}

@article{zhu14,
  title={The large-scale distribution of cool gas around luminous red galaxies},
  author={Zhu, Guangtun and M{\'e}nard, Brice and Bizyaev, Dmitry and Brewington, Howard and Ebelke, Garrett and Ho, Shirley and Kinemuchi, Karen and Malanushenko, Viktor and Malanushenko, Elena and Marchante, Moses and others},
  journal={\mnras},
  volume={439},
  number={3},
  pages={3139--3155},
  year={2014},
  publisher={Oxford University Press},
  doi={10.1093/mnras/stu186},
  archivePrefix={arXiv},
  eprint={1309.7660},
  primaryClass={astro-ph.CO},
  adsurl={https://ui.adsabs.harvard.edu/abs/2014MNRAS.439.3139Z}
}

@article{hahn23,
  title={The desi bright galaxy survey: Final target selection, design, and validation},
  author={Hahn, ChangHoon and Wilson, Michael J and Ruiz-Macias, Omar and Cole, Shaun and Weinberg, David H and Moustakas, John and Kremin, Anthony and Tinker, Jeremy L and Smith, Alex and Wechsler, Risa H and others},
  journal={\aj},
  volume={165},
  number={6},
  pages={253},
  year={2023},
  publisher={IOP Publishing},
  doi={10.3847/1538-3881/accff8},
  archivePrefix={arXiv},
  eprint={2208.08512},
  primaryClass={astro-ph.CO},
  adsurl={https://ui.adsabs.harvard.edu/abs/2023AJ....165..253H}
}

@article{ruiz20,
  title={Preliminary target selection for the DESI bright galaxy survey (BGS)},
  author={Ruiz-Macias, Omar and Zarrouk, Pauline and Cole, Shaun and Norberg, Peder and Baugh, Carlton and Brooks, David and Dey, Arjun and Duan, Yutong and Eftekharzadeh, Sarah and Eisenstein, Daniel J and others},
  journal={Research Notes of the AAS},
  volume={4},
  number={10},
  pages={187},
  year={2020},
  publisher={The American Astronomical Society},
  doi={10.3847/2515-5172/abc25a},
  archivePrefix={arXiv},
  eprint={2010.11283},
  primaryClass={astro-ph.GA},
  adsurl={https://ui.adsabs.harvard.edu/abs/2020RNAAS...4..187R}
}

@article{zhou23,
  title={Target selection and validation of DESI luminous red galaxies},
  author={Zhou, Rongpu and Dey, Biprateep and Newman, Jeffrey A and Eisenstein, Daniel J and Dawson, K and Bailey, S and Berti, A and Guy, J and Lan, Ting-Wen and Zou, H and others},
  journal={\aj},
  volume={165},
  number={2},
  pages={58},
  year={2023},
  publisher={IOP Publishing},
  doi={10.3847/1538-3881/aca5fb},
  archivePrefix={arXiv},
  eprint={2208.08515},
  primaryClass={astro-ph.CO},
  adsurl={https://ui.adsabs.harvard.edu/abs/2023AJ....165...58Z}
}

@article{zhou20,
  title={Preliminary Target Selection for the DESI Luminous Red Galaxy (LRG) Sample},
  author={Zhou, Rongpu and Newman, Jeffrey A and Dawson, Kyle S and Eisenstein, Daniel J and Brooks, David D and Dey, Arjun and Dey, Biprateep and Duan, Yutong and Eftekharzadeh, Sarah and Gazta{\~n}aga, Enrique and others},
  journal={Research Notes of the AAS},
  volume={4},
  number={10},
  pages={181},
  year={2020},
  publisher={The American Astronomical Society},
  doi={10.3847/2515-5172/abc0f4},
  archivePrefix={arXiv},
  eprint={2010.11282},
  primaryClass={astro-ph.CO},
  adsurl={https://ui.adsabs.harvard.edu/abs/2020RNAAS...4..181Z}
}

@article{raichoor23,
  title={Target selection and validation of DESI emission line galaxies},
  author={Raichoor, A and Moustakas, J and Newman, Jeffrey A and Karim, T and Ahlen, S and Alam, Shadab and Bailey, S and Brooks, D and Dawson, K and de la Macorra, A and others},
  journal={\aj},
  volume={165},
  number={3},
  pages={126},
  year={2023},
  publisher={IOP Publishing},
  doi={10.3847/1538-3881/acb213},
  archivePrefix={arXiv},
  eprint={2208.08513},
  primaryClass={astro-ph.CO},
  adsurl={https://ui.adsabs.harvard.edu/abs/2023AJ....165..126R}
}

@article{raichoor20,
  title={Preliminary Target Selection for the DESI Emission Line Galaxy (ELG) Sample},
  author={Raichoor, Anand and Eisenstein, Daniel J and Karim, Tanveer and Newman, Jeffrey A and Moustakas, John and Brooks, David D and Dawson, Kyle S and Dey, Arjun and Duan, Yutong and Eftekharzadeh, Sarah and others},
  journal={Research Notes of the AAS},
  volume={4},
  number={10},
  pages={180},
  year={2020},
  publisher={The American Astronomical Society},
  doi={10.3847/2515-5172/abc078},
  archivePrefix={arXiv},
  eprint={2010.11281},
  primaryClass={astro-ph.CO},
  adsurl={https://ui.adsabs.harvard.edu/abs/2020RNAAS...4..180R}
}

@article{zou24,
  title={A Large Sample of Extremely Metal-poor Galaxies at z< 1 Identified from the DESI Early Data},
  author={Zou, Hu and Sui, Jipeng and Saintonge, Am{\'e}lie and Scholte, Dirk and Moustakas, John and Siudek, Malgorzata and Dey, Arjun and Juneau, Stephanie and Guo, Weijian and Canning, Rebecca and others},
  journal={\apj},
  volume={961},
  number={2},
  pages={173},
  year={2024},
  publisher={IOP Publishing},
  doi={10.3847/1538-4357/ad1409},
  archivePrefix={arXiv},
  eprint={2312.00300},
  primaryClass={astro-ph.GA},
  adsurl={https://ui.adsabs.harvard.edu/abs/2024ApJ...961..173Z}
}

@article{chen10absorb,
  title={An empirical characterization of extended cool gas around galaxies using Mg II absorption features},
  author={Chen, Hsiao-Wen and Helsby, Jennifer E and Gauthier, Jean-Ren{\'e} and Shectman, Stephen A and Thompson, Ian B and Tinker, Jeremy L},
  journal={\apj},
  volume={714},
  number={2},
  pages={1521--1541},
  year={2010},
  publisher={IOP Publishing},
  doi={10.1088/0004-637X/714/2/1521},
  archivePrefix={arXiv},
  eprint={1004.0705},
  primaryClass={astro-ph.CO},
  adsurl={https://ui.adsabs.harvard.edu/abs/2010ApJ...714.1521C}
}

@article{boquien19,
  title={CIGALE: a Python code investigating galaxy emission},
  author={Boquien, M and Burgarella, D and Roehlly, Y and Buat, V and Ciesla, L and Corre, D and Inoue, AK and Salas, H},
  journal={\aap},
  volume={622},
  pages={A103},
  year={2019},
  publisher={EDP Sciences},
  doi={10.1051/0004-6361/201834156},
  archivePrefix={arXiv},
  eprint={1811.03094},
  primaryClass={astro-ph.GA},
  adsurl={https://ui.adsabs.harvard.edu/abs/2019A&A...622A.103B}
}

@article{foreman13,
  title={emcee: the MCMC hammer},
  author={Foreman-Mackey, Daniel and Hogg, David W and Lang, Dustin and Goodman, Jonathan},
  journal={\pasp},
  volume={125},
  number={925},
  pages={306},
  year={2013},
  publisher={IOP Publishing},
  doi={10.1086/670067},
  archivePrefix={arXiv},
  eprint={1202.3665},
  primaryClass={astro-ph.IM},
  adsurl={https://ui.adsabs.harvard.edu/abs/2013PASP..125..306F}
}

@article{napolitano23,
  title={Detecting and Characterizing Mg ii Absorption in DESI Survey Validation Quasar Spectra},
  author={Napolitano, Lucas and Pandey, Agnesh and Myers, Adam D and Lan, Ting-Wen and Anand, Abhijeet and Aguilar, Jessica and Ahlen, Steven and Alexander, David M and Brooks, David and Canning, Rebecca and others},
  journal={\aj},
  volume={166},
  number={3},
  pages={99},
  year={2023},
  publisher={IOP Publishing},
  doi={10.3847/1538-3881/ace62c},
  archivePrefix={arXiv},
  eprint={2305.20016},
  primaryClass={astro-ph.GA},
  adsurl={https://ui.adsabs.harvard.edu/abs/2023AJ....166...99N}
}

@article{chen10cool,
  title={What determines the incidence and extent of Mg II absorbing gas around galaxies?},
  author={Chen, Hsiao-Wen and Wild, Vivienne and Tinker, Jeremy L and Gauthier, Jean-Rene and Helsby, Jennifer E and Shectman, Stephen A and Thompson, Ian B},
  journal={\apjl},
  volume={724},
  number={2},
  pages={L176--L182},
  year={2010},
  publisher={IOP Publishing},
  doi={10.1088/2041-8205/724/2/L176},
  archivePrefix={arXiv},
  eprint={1011.0735},
  primaryClass={astro-ph.CO},
  adsurl={https://ui.adsabs.harvard.edu/abs/2010ApJ...724L.176C}
}

@article{guy2023,
  title={The Spectroscopic Data Processing Pipeline for the Dark Energy Spectroscopic Instrument},
  author={Guy, Julien and Bailey, S and Kremin, A and Alam, Shadab and Alexander, DM and Prieto, C Allende and BenZvi, S and Bolton, AS and Brooks, D and Chaussidon, E and others},
  journal={\aj},
  volume={165},
  number={4},
  pages={144},
  year={2023},
  publisher={IOP Publishing},
  doi={10.3847/1538-3881/acb212},
  archivePrefix={arXiv},
  eprint={2209.14482},
  primaryClass={astro-ph.IM},
  adsurl={https://ui.adsabs.harvard.edu/abs/2023AJ....165..144G}
}

@article{XWChen-21,
  title={A complete census of circumgalactic Mg II at redshift z $\lesssim$ 0.5},
  author={Huang, Yun-Hsin and Chen, Hsiao-Wen and Shectman, Stephen A and Johnson, Sean D and Zahedy, Fakhri S and Helsby, Jennifer E and Gauthier, Jean-Ren{\'e} and Thompson, Ian B},
  journal={\mnras},
  volume={502},
  number={4},
  pages={4743--4761},
  year={2021},
  publisher={Oxford University Press},
  doi={10.1093/mnras/stab360},
  adsurl={https://ui.adsabs.harvard.edu/abs/2021MNRAS.502.4743H}
}

@ARTICLE{Werk2013,
       author = {{Werk}, Jessica K. and {Prochaska}, J. Xavier and {Thom}, Christopher and {Tumlinson}, Jason and {Tripp}, Todd M. and {O'Meara}, John M. and {Peeples}, Molly S.},
        title = "{The COS-Halos Survey: An Empirical Description of Metal-Line Absorption in the Low-redshift Circumgalactic Medium}",
      journal = {\apjs},
         year = 2013,
        month = feb,
       volume = {204},
       number = {2},
          eid = {17},
        pages = {17},
          doi = {10.1088/0067-0049/204/2/17},
archivePrefix = {arXiv},
       eprint = {1212.0558},
 primaryClass = {astro-ph.CO},
       adsurl = {https://ui.adsabs.harvard.edu/abs/2013ApJS..204...17W}
}

@ARTICLE{2016MNRAS.455.1713H,
       author = {{Huang}, Yun-Hsin and {Chen}, Hsiao-Wen and {Johnson}, Sean D. and {Weiner}, Benjamin J.},
        title = "{Characterizing the Chemically Enriched Circumgalactic Medium of $\sim$38,000 Luminous Red Galaxies in SDSS DR12}",
      journal = {\mnras},
         year = 2016,
        month = jan,
       volume = {455},
       number = {2},
        pages = {1713--1727},
          doi = {10.1093/mnras/stv2327},
archivePrefix = {arXiv},
       eprint = {1510.01336},
 primaryClass = {astro-ph.GA},
       adsurl = {https://ui.adsabs.harvard.edu/abs/2016MNRAS.455.1713H}
}

@ARTICLE{2008ApJ...687..745C,
       author = {{Chen}, Hsiao-Wen and {Tinker}, Jeremy L.},
        title = "{The Baryon Content of Dark Matter Halos: Empirical Constraints from Mg II Absorbers}",
      journal = {\apj},
         year = 2008,
        month = nov,
       volume = {687},
       number = {2},
        pages = {745--756},
          doi = {10.1086/591927},
archivePrefix = {arXiv},
       eprint = {0801.2169},
 primaryClass = {astro-ph},
       adsurl = {https://ui.adsabs.harvard.edu/abs/2008ApJ...687..745C}
}

@INPROCEEDINGS{2026enap....4..370C,
       author = {{Chen}, Hsiao-Wen and {Zahedy}, Fakhri S.},
        title = "{The Circumgalactic Medium}",
    booktitle = {Encyclopedia of Astrophysics, Volume 4},
         year = 2026,
       volume = {4},
        month = jan,
        pages = {370--400},
          doi = {10.1016/B978-0-443-21439-4.00059-6},
archivePrefix = {arXiv},
       eprint = {2412.10579},
 primaryClass = {astro-ph.GA},
       adsurl = {https://ui.adsabs.harvard.edu/abs/2026enap....4..370C}
}

@ARTICLE{churchill2013,
       author = {{Churchill}, Christopher W. and {Trujillo-Gomez}, Sebastian and {Nielsen}, Nikole M. and {Kacprzak}, Glenn G.},
        title = "{MAGIICAT III. Interpreting Self-Similarity of the Circumgalactic Medium with Virial Mass Using Mg II Absorption}",
      journal = {\apj},
         year = 2013,
        month = dec,
       volume = {779},
       number = {1},
          eid = {87},
        pages = {87},
          doi = {10.1088/0004-637X/779/1/87},
archivePrefix = {arXiv},
       eprint = {1308.2618},
 primaryClass = {astro-ph.CO},
       adsurl = {https://ui.adsabs.harvard.edu/abs/2013ApJ...779...87C}
}

@ARTICLE{nielsen2016,
       author = {{Nielsen}, Nikole M. and {Churchill}, Christopher W. and {Kacprzak}, Glenn G. and {Murphy}, Michael T. and {Evans}, Jessica L.},
        title = "{MAGIICAT IV. Kinematics of the Circumgalactic Medium and Evidence for Quiescent Evolution Around Red Galaxies}",
      journal = {\apj},
         year = 2016,
        month = feb,
       volume = {818},
       number = {2},
          eid = {171},
        pages = {171},
          doi = {10.3847/0004-637X/818/2/171},
archivePrefix = {arXiv},
       eprint = {1512.06969},
 primaryClass = {astro-ph.GA},
       adsurl = {https://ui.adsabs.harvard.edu/abs/2016ApJ...818..171N}
}

@article{wu2024,
  title={Tracing the evolution of the cool gas in CGM and IGM environments through Mg II absorption from redshift z= 0.75 to z= 1.65 using DESI-Y1 data},
  author={Wu, Xuanyi and Cai, Z and Lan, T-W and Zou, S and Anand, A and Dey, Biprateep and Li, Z and Aguilar, J and Ahlen, S and Brooks, D and others},
  journal={\apj},
  volume={983},
  number={2},
  pages={186},
  doi={10.3847/1538-4357/adb28a},
  year={2025},
  publisher={IOP Publishing},
  adsurl={https://ui.adsabs.harvard.edu/abs/2025ApJ...983..186W}
}

@article{chen25,
  title={The Circumgalactic Medium Traced by Mg ii Absorption with DESI: Dependence on Galaxy Stellar Mass, Star Formation Rate, and Azimuthal Angle},
  author={Chen, Zeyu and Wang, Enci and Zou, Hu and Zou, Siwei and Gao, Yang and Wang, Huiyuan and Yu, Haoran and Jia, Cheng and Li, Haixin and Ma, Chengyu and others},
  journal={\apj},
  volume={981},
  number={1},
  pages={81},
  year={2025a},
  doi={10.3847/1538-4357/ada942},
  publisher={IOP Publishing},
  archivePrefix={arXiv},
  eprint={2411.08485},
  primaryClass={astro-ph.GA},
  adsurl={https://ui.adsabs.harvard.edu/abs/2025ApJ...981...81C}
}

@ARTICLE{DESI2025,
       author = {{DESI Collaboration} and {Karim}, M. Abdul and {Adame}, A.~G. and {Aguado}, D. and {Aguilar}, J. and {Ahlen}, S. and {Alam}, S. and {Aldering}, G. and {Alexander}, D.~M. and {Alfarsy}, R. and {Allen}, L. and {Allende Prieto}, C. and {Alves}, O. and {Anand}, A. and {Andrade}, U. and {Armengaud}, E. and {Avila}, S. and {Aviles}, A. and {Awan}, H. and {Bailey}, S. and {Baleato Lizancos}, A. and {Ballester}, O. and {Bault}, A. and {Bautista}, J. and {BenZvi}, S. and {Beraldo e Silva}, L. and {Bermejo-Climent}, J.~R. and {Beutler}, F. and {Bianchi}, D. and {Blake}, C. and {Blum}, R. and {Bolton}, A.~S. and {Bonici}, M. and {Brieden}, S. and {Brodzeller}, A. and {Brooks}, D. and {Buckley-Geer}, E. and {Burtin}, E. and {Canning}, R. and {Carnero Rosell}, A. and {Carr}, A. and {Carrilho}, P. and {Casas}, L. and {Castander}, F.~J. and {Cereskaite}, R. and {Cervantes-Cota}, J.~L. and {Chaussidon}, E. and {Chaves-Montero}, J. and {Chen}, S. and {Chen}, X. and {Claybaugh}, T. and {Cole}, S. and {Cooper}, A.~P. and {Cousinou}, M. -C. and {Cuceu}, A. and {Davis}, T.~M. and {Dawson}, K.~S. and {de Belsunce}, R. and {de la Cruz}, R. and {de la Macorra}, A. and {de Mattia}, A. and {Deiosso}, N. and {Della Costa}, J. and {Demina}, R. and {Demirbozan}, U. and {DeRose}, J. and {Dey}, A. and {Dey}, B. and {Ding}, J. and {Ding}, Z. and {Doel}, P. and {Douglass}, K. and {Dowicz}, M. and {Ebina}, H. and {Edelstein}, J. and {Eisenstein}, D.~J. and {Elbers}, W. and {Emas}, N. and {Escoffier}, S. and {Fagrelius}, P. and {Fan}, X. and {Fanning}, K. and {Fawcett}, V.~A. and {Fern{\'a}ndez-Garc{\'\i}a}, E. and {Ferraro}, S. and {Findlay}, N. and {Font-Ribera}, A. and {Forero-Romero}, J.~E. and {Forero-S{\'a}nchez}, D. and {Frenk}, C.~S. and {G{\"a}nsicke}, B.~T. and {Galbany}, L. and {Garc{\'\i}a-Bellido}, J. and {Garcia-Quintero}, C. and {Garrison}, L.~H. and {Gazta{\~n}aga}, E. and {Gil-Mar{\'\i}n}, H. and {Gnedin}, O.~Y. and {Gontcho}, S. Gontcho A and {Gonzalez-Morales}, A.~X. and {Gonzalez-Perez}, V. and {Gordon}, C. and {Graur}, O. and {Green}, D. and {Gruen}, D. and {Gsponer}, R. and {Guandalin}, C. and {Gutierrez}, G. and {Guy}, J. and {Hahn}, C. and {Han}, J.~J. and {Han}, J. and {He}, S. and {Herrera-Alcantar}, H.~K. and {Honscheid}, K. and {Hou}, J. and {Howlett}, C. and {Huterer}, D. and {Ir{\v{s}}i{\v{c}}}, V. and {Ishak}, M. and {Jacques}, A. and {Jimenez}, J. and {Jing}, Y.~P. and {Joachimi}, B. and {Joudaki}, S. and {Joyce}, R. and {Jullo}, E. and {Juneau}, S. and {Kara{\c{c}}ayl{\i}}, N.~G. and {Karim}, T. and {Kehoe}, R. and {Kent}, S. and {Khederlarian}, A. and {Kirkby}, D. and {Kisner}, T. and {Kitaura}, F. -S. and {Kizhuprakkat}, N. and {Kong}, H. and {Koposov}, S.~E. and {Kremin}, A. and {Krolewski}, A. and {Lahav}, O. and {Lai}, Y. and {Lamman}, C. and {Lan}, T. -W. and {Landriau}, M. and {Lang}, D. and {Lange}, J.~U. and {Lasker}, J. and {Le Goff}, J.~M. and {Le Guillou}, L. and {Leauthaud}, A. and {Levi}, M.~E. and {Li}, S. and {Li}, T.~S. and {Lodha}, K. and {Lokken}, M. and {Luo}, Y. and {Magneville}, C. and {Manera}, M. and {Manser}, C.~J. and {Margala}, D. and {Martini}, P. and {Maus}, M. and {McCullough}, J. and {McDonald}, P. and {Medina}, G.~E. and {Medina-Varela}, L. and {Meisner}, A. and {Mena-Fern{\'a}ndez}, J. and {Menegas}, A. and {Mezcua}, M. and {Miquel}, R. and {Montero-Camacho}, P. and {Moon}, J. and {Moustakas}, J. and {Mu{\~n}oz-Guti{\'e}rrez}, A. and {Mu{\~n}oz-Santos}, D. and {Myers}, A.~D. and {Myles}, J. and {Nadathur}, S. and {Najita}, J. and {Napolitano}, L. and {Newman}, J.~A. and {Nikakhtar}, F. and {Nikutta}, R. and {Niz}, G. and {Noriega}, H.~E. and {Padmanabhan}, N. and {Paillas}, E. and {Palanque-Delabrouille}, N. and {Palmese}, A. and {Pan}, J. and {Pan}, Z. and {Parkinson}, D. and {Peacock}, J. and {Percival}, W.~J. and {P{\'e}rez-Fern{\'a}ndez}, A. and {P{\'e}rez-R{\`a}fols}, I. and {Peterson}, P.},
        title = "{Data Release 1 of the Dark Energy Spectroscopic Instrument}",
      journal = {arXiv e-prints},
         year = {2025},
        month = mar,
          eid = {arXiv:2503.14745},
        pages = {arXiv:2503.14745},
          doi = {10.48550/arXiv.2503.14745},
archivePrefix = {arXiv},
       eprint = {2503.14745},
 primaryClass = {astro-ph.CO},
       adsurl = {https://ui.adsabs.harvard.edu/abs/2025arXiv250314745D}
}

@ARTICLE{levi13,
       author = {{Levi}, Michael and {Bebek}, Chris and {Beers}, Timothy and {Blum}, Robert and {Cahn}, Robert and {Eisenstein}, Daniel and {Flaugher}, Brenna and {Honscheid}, Klaus and {Kron}, Richard and {Lahav}, Ofer and {McDonald}, Patrick and {Roe}, Natalie and {Schlegel}, David and {representing the DESI collaboration}},
        title = "{The DESI Experiment, a whitepaper for Snowmass 2013}",
      journal = {arXiv e-prints},
         year = 2013,
        month = aug,
          eid = {arXiv:1308.0847},
        pages = {arXiv:1308.0847},
          doi = {10.48550/arXiv.1308.0847},
archivePrefix = {arXiv},
       eprint = {1308.0847},
 primaryClass = {astro-ph.CO},
       adsurl = {https://ui.adsabs.harvard.edu/abs/2013arXiv1308.0847L}
}

@misc{lang16,
       author = {{Lang}, Dustin and {Hogg}, David W. and {Mykytyn}, David},
        title = "{The Tractor: Probabilistic astronomical source detection and measurement}",
 howpublished = {Astrophysics Source Code Library, record ascl:1604.008},
         year = 2016,
        month = apr,
          eid = {ascl:1604.008},
       adsurl = {https://ui.adsabs.harvard.edu/abs/2016ascl.soft04008L}
}

@ARTICLE{anand24,
       author = {{Anand}, Abhijeet and {Guy}, Julien and {Bailey}, Stephen and {Moustakas}, John and {Aguilar}, J. and {Ahlen}, S. and {Bolton}, A.~S. and {Brodzeller}, A. and {Brooks}, D. and {Claybaugh}, T. and {Cole}, S. and {de la Macorra}, A. and {Dey}, Biprateep and {Fanning}, K. and {Forero-Romero}, J.~E. and {Gazta{\~n}aga}, E. and {Gontcho A Gontcho}, S. and {Gutierrez}, G. and {Honscheid}, K. and {Howlett}, C. and {Juneau}, S. and {Kirkby}, D. and {Kisner}, T. and {Kremin}, A. and {Lambert}, A. and {Landriau}, M. and {Le Guillou}, L. and {Manera}, M. and {Meisner}, A. and {Miquel}, R. and {Mueller}, E. and {Niz}, G. and {Palanque-Delabrouille}, N. and {Percival}, W.~J. and {Poppett}, C. and {Prada}, F. and {Raichoor}, A. and {Rezaie}, M. and {Rossi}, G. and {Sanchez}, E. and {Schlafly}, E.~F. and {Schlegel}, D. and {Schubnell}, M. and {Sprayberry}, D. and {Tarl{\'e}}, G. and {Warner}, C. and {Weaver}, B.~A. and {Zhou}, R. and {Zou}, H.},
        title = "{Archetype-based Redshift Estimation for the Dark Energy Spectroscopic Instrument Survey}",
      journal = {\aj},
         year = 2024,
        month = sep,
       volume = {168},
       number = {3},
          eid = {124},
        pages = {124},
          doi = {10.3847/1538-3881/ad60c2},
archivePrefix = {arXiv},
       eprint = {2405.19288},
 primaryClass = {astro-ph.CO},
       adsurl = {https://ui.adsabs.harvard.edu/abs/2024AJ....168..124A}
}

@article{ng2025,
  title={A Comprehensive Characterization of Galaxy-Cool CGM Connections at $ z< 0.4$ with DESI Year 1 Data},
  author={Ng, Yu Voon and Lan, Ting-Wen and Prochaska, J Xavier and Saintonge, Am{\'e}lie and Chang, Yu-Ling and Siudek, Ma{\l}gorzata and Aguilar, Jessica Nicole and Ahlen, Steven and Bianchi, Davide and Brooks, David and others},
  journal={\apj},
  year={2025},
  volume={993},
  number={1},
  eid={92},
  pages={92},
  doi={10.3847/1538-4357/ae0613},
  archivePrefix={arXiv},
  eprint={2503.11139},
  primaryClass={astro-ph.GA},
  adsurl={https://ui.adsabs.harvard.edu/abs/2025ApJ...993...92N}
}

@ARTICLE{siudek24,
       author = {{Siudek}, M. and {Pucha}, R. and {Mezcua}, M. and {Juneau}, S. and {Aguilar}, J. and {Ahlen}, S. and {Brooks}, D. and {Circosta}, C. and {Claybaugh}, T. and {Cole}, S. and {Dawson}, K. and {de la Macorra}, A. and {Dey}, A. and {Dey}, B. and {Doel}, P. and {Font-Ribera}, A. and {Forero-Romero}, J.~E. and {Gazta{\~n}aga}, E. and {Gontcho A Gontcho}, S. and {Gutierrez}, G. and {Honscheid}, K. and {Howlett}, C. and {Ishak}, M. and {Kehoe}, R. and {Kirkby}, D. and {Kisner}, T. and {Kremin}, A. and {Lambert}, A. and {Landriau}, M. and {Le Guillou}, L. and {Manera}, M. and {Martini}, P. and {Meisner}, A. and {Miquel}, R. and {Moustakas}, J. and {Newman}, J.~A. and {Niz}, G. and {Pan}, Z. and {Percival}, W.~J. and {Poppett}, C. and {Prada}, F. and {Rossi}, G. and {Saintonge}, A. and {Sanchez}, E. and {Schlegel}, D. and {Scholte}, D. and {Schubnell}, M. and {Seo}, H. and {Speranza}, F. and {Sprayberry}, D. and {Tarl{\'e}}, G. and {Weaver}, B.~A. and {Zou}, H.},
        title = "{Value-added catalog of physical properties for more than 1.3 million galaxies from the DESI survey}",
      journal = {\aap},
         year = 2024,
        month = nov,
       volume = {691},
          eid = {A308},
        pages = {A308},
          doi = {10.1051/0004-6361/202451761},
archivePrefix = {arXiv},
       eprint = {2409.19066},
 primaryClass = {astro-ph.GA},
       adsurl = {https://ui.adsabs.harvard.edu/abs/2024A&A...691A.308S}
}

@ARTICLE{moustakas13,
       author = {{Moustakas}, John and {Coil}, Alison L. and {Aird}, James and {Blanton}, Michael R. and {Cool}, Richard J. and {Eisenstein}, Daniel J. and {Mendez}, Alexander J. and {Wong}, Kenneth C. and {Zhu}, Guangtun and {Arnouts}, St{\'e}phane},
        title = "{PRIMUS: Constraints on Star Formation Quenching and Galaxy Merging, and the Evolution of the Stellar Mass Function from z = 0-1}",
      journal = {\apj},
         year = 2013,
        month = apr,
       volume = {767},
       number = {1},
          eid = {50},
        pages = {50},
          doi = {10.1088/0004-637X/767/1/50},
archivePrefix = {arXiv},
       eprint = {1301.1688},
 primaryClass = {astro-ph.CO},
       adsurl = {https://ui.adsabs.harvard.edu/abs/2013ApJ...767...50M}
}

@ARTICLE{nelson19,
       author = {{Nelson}, Dylan and {Pillepich}, Annalisa and {Springel}, Volker and {Pakmor}, R{\"u}diger and {Weinberger}, Rainer and {Genel}, Shy and {Torrey}, Paul and {Vogelsberger}, Mark and {Marinacci}, Federico and {Hernquist}, Lars},
        title = "{First results from the TNG50 simulation: galactic outflows driven by supernovae and black hole feedback}",
      journal = {\mnras},
         year = 2019,
        month = dec,
       volume = {490},
       number = {3},
        pages = {3234-3261},
          doi = {10.1093/mnras/stz2306},
archivePrefix = {arXiv},
       eprint = {1902.05554},
 primaryClass = {astro-ph.GA},
       adsurl = {https://ui.adsabs.harvard.edu/abs/2019MNRAS.490.3234N}
}

@ARTICLE{wright21,
       author = {{Wright}, Ruby J. and {Lagos}, Claudia del P. and {Power}, Chris and {Correa}, Camila A.},
        title = "{Revealing the physical properties of gas accreting to haloes in the EAGLE simulations}",
      journal = {\mnras},
         year = 2021,
        month = jul,
       volume = {504},
       number = {4},
        pages = {5702-5725},
          doi = {10.1093/mnras/stab1057},
archivePrefix = {arXiv},
       eprint = {2102.10913},
 primaryClass = {astro-ph.GA},
       adsurl = {https://ui.adsabs.harvard.edu/abs/2021MNRAS.504.5702W}
}

@ARTICLE{schaye15,
       author = {{Schaye}, Joop and {Crain}, Robert A. and {Bower}, Richard G. and {Furlong}, Michelle and {Schaller}, Matthieu and {Theuns}, Tom and {Dalla Vecchia}, Claudio and {Frenk}, Carlos S. and {McCarthy}, I.~G. and {Helly}, John C. and {Jenkins}, Adrian and {Rosas-Guevara}, Y.~M. and {White}, Simon D.~M. and {Baes}, Maarten and {Booth}, C.~M. and {Camps}, Peter and {Navarro}, Julio F. and {Qu}, Yan and {Rahmati}, Alireza and {Sawala}, Till and {Thomas}, Peter A. and {Trayford}, James},
        title = "{The EAGLE project: simulating the evolution and assembly of galaxies and their environments}",
      journal = {\mnras},
         year = 2015,
        month = jan,
       volume = {446},
       number = {1},
        pages = {521-554},
          doi = {10.1093/mnras/stu2058},
archivePrefix = {arXiv},
       eprint = {1407.7040},
 primaryClass = {astro-ph.GA},
       adsurl = {https://ui.adsabs.harvard.edu/abs/2015MNRAS.446..521S}
}

@ARTICLE{hafen20,
       author = {{Hafen}, Zachary and {Faucher-Gigu{\`e}re}, Claude-Andr{\'e} and {Angl{\'e}s-Alc{\'a}zar}, Daniel and {Stern}, Jonathan and {Kere{\v{s}}}, Du{\v{s}}an and {Esmerian}, Clarke and {Wetzel}, Andrew and {El-Badry}, Kareem and {Chan}, T.~K. and {Murray}, Norman},
        title = "{The fates of the circumgalactic medium in the FIRE simulations}",
      journal = {\mnras},
         year = 2020,
        month = may,
       volume = {494},
       number = {3},
        pages = {3581-3595},
          doi = {10.1093/mnras/staa902},
archivePrefix = {arXiv},
       eprint = {1910.01123},
 primaryClass = {astro-ph.GA},
       adsurl = {https://ui.adsabs.harvard.edu/abs/2020MNRAS.494.3581H}
}

@ARTICLE{moster18,
       author = {{Moster}, Benjamin P. and {Naab}, Thorsten and {White}, Simon D.~M.},
        title = "{EMERGE - an empirical model for the formation of galaxies since z {\ensuremath{\sim}} 10}",
      journal = {\mnras},
         year = 2018,
        month = jun,
       volume = {477},
       number = {2},
        pages = {1822-1852},
          doi = {10.1093/mnras/sty655},
archivePrefix = {arXiv},
       eprint = {1705.05373},
 primaryClass = {astro-ph.GA},
       adsurl = {https://ui.adsabs.harvard.edu/abs/2018MNRAS.477.1822M}
}

@ARTICLE{moster13,
       author = {{Moster}, Benjamin P. and {Naab}, Thorsten and {White}, Simon D.~M.},
        title = "{Galactic star formation and accretion histories from matching galaxies to dark matter haloes}",
      journal = {\mnras},
         year = 2013,
        month = feb,
       volume = {428},
       number = {4},
        pages = {3121-3138},
          doi = {10.1093/mnras/sts261},
archivePrefix = {arXiv},
       eprint = {1205.5807},
 primaryClass = {astro-ph.CO},
       adsurl = {https://ui.adsabs.harvard.edu/abs/2013MNRAS.428.3121M}
}

@ARTICLE{girelli20,
       author = {{Girelli}, G. and {Pozzetti}, L. and {Bolzonella}, M. and {Giocoli}, C. and {Marulli}, F. and {Baldi}, M.},
        title = "{The stellar-to-halo mass relation over the past 12 Gyr. I. Standard {\ensuremath{\Lambda}}CDM model}",
      journal = {\aap},
         year = 2020,
        month = feb,
       volume = {634},
          eid = {A135},
        pages = {A135},
          doi = {10.1051/0004-6361/201936329},
archivePrefix = {arXiv},
       eprint = {2001.02230},
 primaryClass = {astro-ph.CO},
       adsurl = {https://ui.adsabs.harvard.edu/abs/2020A&A...634A.135G}
}

@ARTICLE{shuntov22,
       author = {{Shuntov}, M. and {McCracken}, H.~J. and {Gavazzi}, R. and {Laigle}, C. and {Weaver}, J.~R. and {Davidzon}, I. and {Ilbert}, O. and {Kauffmann}, O.~B. and {Faisst}, A. and {Dubois}, Y. and {Koekemoer}, A.~M. and {Moneti}, A. and {Milvang-Jensen}, B. and {Mobasher}, B. and {Sanders}, D.~B. and {Toft}, S.},
        title = "{COSMOS2020: Cosmic evolution of the stellar-to-halo mass relation for central and satellite galaxies up to z {\ensuremath{\sim}} 5}",
      journal = {\aap},
         year = 2022,
        month = aug,
       volume = {664},
          eid = {A61},
        pages = {A61},
          doi = {10.1051/0004-6361/202243136},
archivePrefix = {arXiv},
       eprint = {2203.10895},
 primaryClass = {astro-ph.GA},
       adsurl = {https://ui.adsabs.harvard.edu/abs/2022A&A...664A..61S}
}

@BOOK{mo10,
       author = {{Mo}, Houjun and {van den Bosch}, Frank C. and {White}, Simon},
        title = "{Galaxy Formation and Evolution}",
         year = 2010,
          doi = {10.1017/CBO9780511807244},
       adsurl = {https://ui.adsabs.harvard.edu/abs/2010gfe..book.....M},
    publisher = {Cambridge University Press}
}

@ARTICLE{legrand19,
       author = {{Legrand}, L. and {McCracken}, H.~J. and {Davidzon}, I. and {Ilbert}, O. and {Coupon}, J. and {Aghanim}, N. and {Douspis}, M. and {Capak}, P.~L. and {Le F{\`e}vre}, O. and {Milvang-Jensen}, B.},
        title = "{The COSMOS-UltraVISTA stellar-to-halo mass relationship: new insights on galaxy formation efficiency out to z {\ensuremath{\sim}} 5}",
      journal = {\mnras},
         year = 2019,
        month = jul,
       volume = {486},
       number = {4},
        pages = {5468-5481},
          doi = {10.1093/mnras/stz1198},
archivePrefix = {arXiv},
       eprint = {1810.10557},
 primaryClass = {astro-ph.GA},
       adsurl = {https://ui.adsabs.harvard.edu/abs/2019MNRAS.486.5468L}
}

@ARTICLE{kravtsov18,
       author = {{Kravtsov}, A.~V. and {Vikhlinin}, A.~A. and {Meshcheryakov}, A.~V.},
        title = "{Stellar Mass{\textemdash}Halo Mass Relation and Star Formation Efficiency in High-Mass Halos}",
      journal = {Astronomy Letters},
         year = 2018,
        month = jan,
       volume = {44},
       number = {1},
        pages = {8-34},
          doi = {10.1134/S1063773717120015},
archivePrefix = {arXiv},
       eprint = {1401.7329},
 primaryClass = {astro-ph.CO},
       adsurl = {https://ui.adsabs.harvard.edu/abs/2018AstL...44....8K}
}

@book{dopita03,
  title={Astrophysics of the Diffuse Universe},
  author={Dopita, Michael A and Sutherland, Ralph S},
  year={2003},
  publisher={Springer},
  address={Berlin, Heidelberg},
  isbn={978-3-540-43362-0},
  doi={10.1007/978-3-662-05866-4}
}

@ARTICLE{bilicki21,
       author = {{Bilicki}, M. and {Dvornik}, A. and {Hoekstra}, H. and {Wright}, A.~H. and {Chisari}, N.~E. and {Vakili}, M. and {Asgari}, M. and {Giblin}, B. and {Heymans}, C. and {Hildebrandt}, H. and {Holwerda}, B.~W. and {Hopkins}, A. and {Johnston}, H. and {Kannawadi}, A. and {Kuijken}, K. and {Nakoneczny}, S.~J. and {Shan}, H.~Y. and {Sonnenfeld}, A. and {Valentijn}, E.},
        title = "{Bright galaxy sample in the Kilo-Degree Survey Data Release 4. Selection, photometric redshifts, and physical properties}",
      journal = {\aap},
         year = 2021,
        month = sep,
       volume = {653},
          eid = {A82},
        pages = {A82},
          doi = {10.1051/0004-6361/202140352},
archivePrefix = {arXiv},
       eprint = {2101.06010},
 primaryClass = {astro-ph.GA},
       adsurl = {https://ui.adsabs.harvard.edu/abs/2021A&A...653A..82B}
}

@article{wang2025,
  title={Testing Galaxy Formation Models with the Stellar Mass--Halo Mass Relations for Star-forming and Quiescent Galaxies},
  author={Wang, Kai and Peng, Yingjie},
  journal={\apj},
  volume={980},
  number={2},
  pages={233},
  year={2025},
  doi={10.3847/1538-4357/adaf1a},
  publisher={IOP Publishing},
  adsurl={https://ui.adsabs.harvard.edu/abs/2025ApJ...980..233W}
}

@ARTICLE{chen2025b,
       author = {{Chen}, Zeyu and {Wang}, Enci and {Zou}, Hu and {Yu}, Haoran and {He}, Zhicheng and {Wang}, Huiyuan and {Gao}, Yang and {Lyu}, Cheqiu and {Jia}, Cheng and {Ma}, Chengyu and {Ding}, Weiyu and {Zhu}, Runyu and {Kong}, Xu},
        title = "{The Cosmic Evolution and Spatial Distribution of Multiphase Gas Associated with QSOs}",
      journal = {\apjl},
         year = {2025b},
        month = jul,
       volume = {988},
       number = {1},
          eid = {L39},
        pages = {L39},
          doi = {10.3847/2041-8213/ade545},
archivePrefix = {arXiv},
       eprint = {2505.11919},
 primaryClass = {astro-ph.GA},
       adsurl = {https://ui.adsabs.harvard.edu/abs/2025ApJ...988L..39C}
}

@ARTICLE{McGaugh14,
       author = {{McGaugh}, Stacy S. and {Schombert}, James M.},
        title = "{Color-Mass-to-light-ratio Relations for Disk Galaxies}",
      journal = {\aj},
         year = 2014,
        month = nov,
       volume = {148},
       number = {5},
          eid = {77},
        pages = {77},
          doi = {10.1088/0004-6256/148/5/77},
archivePrefix = {arXiv},
       eprint = {1407.1839},
 primaryClass = {astro-ph.GA},
       adsurl = {https://ui.adsabs.harvard.edu/abs/2014AJ....148...77M}
}

@ARTICLE{Mandelbaum16,
       author = {{Mandelbaum}, Rachel and {Wang}, Wenting and {Zu}, Ying and {White}, Simon and {Henriques}, Bruno and {More}, Surhud},
        title = "{Strong bimodality in the host halo mass of central galaxies from galaxy-galaxy lensing}",
      journal = {\mnras},
         year = 2016,
        month = apr,
       volume = {457},
       number = {3},
        pages = {3200-3218},
          doi = {10.1093/mnras/stw188},
archivePrefix = {arXiv},
       eprint = {1509.06762},
 primaryClass = {astro-ph.GA},
       adsurl = {https://ui.adsabs.harvard.edu/abs/2016MNRAS.457.3200M}
}

@ARTICLE{zhang22,
       author = {{Zhang}, Ziwen and {Wang}, Huiyuan and {Luo}, Wentao and {Zhang}, Jun and {Mo}, Houjun J. and {Jing}, YiPeng and {Yang}, Xiaohu and {Li}, Hao},
        title = "{Massive star-forming galaxies have converted most of their halo gas into stars}",
      journal = {\aap},
         year = 2022,
        month = jul,
       volume = {663},
          eid = {A85},
        pages = {A85},
          doi = {10.1051/0004-6361/202142866},
archivePrefix = {arXiv},
       eprint = {2112.04777},
 primaryClass = {astro-ph.GA},
       adsurl = {https://ui.adsabs.harvard.edu/abs/2022A&A...663A..85Z}
}

@ARTICLE{moster10,
       author = {{Moster}, Benjamin P. and {Somerville}, Rachel S. and {Maulbetsch}, Christian and {van den Bosch}, Frank C. and {Macci{\`o}}, Andrea V. and {Naab}, Thorsten and {Oser}, Ludwig},
        title = "{Constraints on the Relationship between Stellar Mass and Halo Mass at Low and High Redshift}",
      journal = {\apj},
         year = 2010,
        month = feb,
       volume = {710},
       number = {2},
        pages = {903-923},
          doi = {10.1088/0004-637X/710/2/903},
archivePrefix = {arXiv},
       eprint = {0903.4682},
 primaryClass = {astro-ph.CO},
       adsurl = {https://ui.adsabs.harvard.edu/abs/2010ApJ...710..903M}
}

@ARTICLE{rodr17,
       author = {{Rodr{\'\i}guez-Puebla}, Aldo and {Primack}, Joel R. and {Avila-Reese}, Vladimir and {Faber}, S.~M.},
        title = "{Constraining the galaxy-halo connection over the last 13.3 Gyr: star formation histories, galaxy mergers and structural properties}",
      journal = {\mnras},
         year = 2017,
        month = sep,
       volume = {470},
       number = {1},
        pages = {651-687},
          doi = {10.1093/mnras/stx1172},
archivePrefix = {arXiv},
       eprint = {1703.04542},
 primaryClass = {astro-ph.GA},
       adsurl = {https://ui.adsabs.harvard.edu/abs/2017MNRAS.470..651R}
}

@ARTICLE{Dekel06,
       author = {{Dekel}, Avishai and {Birnboim}, Yuval},
        title = "{Galaxy bimodality due to cold flows and shock heating}",
      journal = {\mnras},
         year = 2006,
        month = may,
       volume = {368},
       number = {1},
        pages = {2-20},
          doi = {10.1111/j.1365-2966.2006.10145.x},
archivePrefix = {arXiv},
       eprint = {astro-ph/0412300},
 primaryClass = {astro-ph},
       adsurl = {https://ui.adsabs.harvard.edu/abs/2006MNRAS.368....2D}
}

@ARTICLE{Vogelsberger20,
       author = {{Vogelsberger}, Mark and {Marinacci}, Federico and {Torrey}, Paul and {Puchwein}, Ewald},
        title = "{Cosmological simulations of galaxy formation}",
      journal = {Nature Reviews Physics},
         year = 2020,
        month = jan,
       volume = {2},
       number = {1},
        pages = {42-66},
          doi = {10.1038/s42254-019-0127-2},
archivePrefix = {arXiv},
       eprint = {1909.07976},
 primaryClass = {astro-ph.GA},
       adsurl = {https://ui.adsabs.harvard.edu/abs/2020NatRP...2...42V}
}

@ARTICLE{Mitchell22,
       author = {{Mitchell}, Peter D. and {Schaye}, Joop},
        title = "{How gas flows shape the stellar-halo mass relation in the EAGLE simulation}",
      journal = {\mnras},
         year = 2022,
        month = apr,
       volume = {511},
       number = {2},
        pages = {2948-2967},
          doi = {10.1093/mnras/stab3339},
archivePrefix = {arXiv},
       eprint = {2103.10966},
 primaryClass = {astro-ph.GA},
       adsurl = {https://ui.adsabs.harvard.edu/abs/2022MNRAS.511.2948M}
}

@ARTICLE{Rees1977,
       author = {{Rees}, M.~J. and {Ostriker}, J.~P.},
        title = "{Cooling, dynamics and fragmentation of massive gas clouds: clues to the masses and radii of galaxies and clusters.}",
      journal = {\mnras},
         year = 1977,
        month = jun,
       volume = {179},
        pages = {541-559},
          doi = {10.1093/mnras/179.4.541},
       adsurl = {https://ui.adsabs.harvard.edu/abs/1977MNRAS.179..541R}
}

@ARTICLE{White1978,
       author = {{White}, S.~D.~M. and {Rees}, M.~J.},
        title = "{Core condensation in heavy halos: a two-stage theory for galaxy formation and clustering.}",
      journal = {\mnras},
         year = 1978,
        month = may,
       volume = {183},
        pages = {341-358},
          doi = {10.1093/mnras/183.3.341},
       adsurl = {https://ui.adsabs.harvard.edu/abs/1978MNRAS.183..341W}
}

@ARTICLE{bower06,
       author = {{Bower}, R.~G. and {Benson}, A.~J. and {Malbon}, R. and {Helly}, J.~C. and {Frenk}, C.~S. and {Baugh}, C.~M. and {Cole}, S. and {Lacey}, C.~G.},
        title = "{Breaking the hierarchy of galaxy formation}",
      journal = {\mnras},
         year = 2006,
        month = aug,
       volume = {370},
       number = {2},
        pages = {645-655},
          doi = {10.1111/j.1365-2966.2006.10519.x},
archivePrefix = {arXiv},
       eprint = {astro-ph/0511338},
 primaryClass = {astro-ph},
       adsurl = {https://ui.adsabs.harvard.edu/abs/2006MNRAS.370..645B}
}

@ARTICLE{croton06,
       author = {{Croton}, Darren J. and {Springel}, Volker and {White}, Simon D.~M. and {De Lucia}, G. and {Frenk}, C.~S. and {Gao}, L. and {Jenkins}, A. and {Kauffmann}, G. and {Navarro}, J.~F. and {Yoshida}, N.},
        title = "{The many lives of active galactic nuclei: cooling flows, black holes and the luminosities and colours of galaxies}",
      journal = {\mnras},
         year = 2006,
        month = jan,
       volume = {365},
       number = {1},
        pages = {11-28},
          doi = {10.1111/j.1365-2966.2005.09675.x},
archivePrefix = {arXiv},
       eprint = {astro-ph/0508046},
 primaryClass = {astro-ph},
       adsurl = {https://ui.adsabs.harvard.edu/abs/2006MNRAS.365...11C}
}

@ARTICLE{Wechsler18,
       author = {{Wechsler}, Risa H. and {Tinker}, Jeremy L.},
        title = "{The Connection Between Galaxies and Their Dark Matter Halos}",
      journal = {\araa},
         year = 2018,
        month = sep,
       volume = {56},
        pages = {435-487},
          doi = {10.1146/annurev-astro-081817-051756},
archivePrefix = {arXiv},
       eprint = {1804.03097},
 primaryClass = {astro-ph.GA},
       adsurl = {https://ui.adsabs.harvard.edu/abs/2018ARA&A..56..435W}
}

@ARTICLE{Posti21,
       author = {{Posti}, L. and {Fall}, S.~M.},
        title = "{Dynamical evidence for a morphology-dependent relation between the stellar and halo masses of galaxies}",
      journal = {\aap},
         year = 2021,
        month = may,
       volume = {649},
          eid = {A119},
        pages = {A119},
          doi = {10.1051/0004-6361/202040256},
archivePrefix = {arXiv},
       eprint = {2102.11282},
 primaryClass = {astro-ph.GA},
       adsurl = {https://ui.adsabs.harvard.edu/abs/2021A&A...649A.119P}
}

@ARTICLE{more11,
       author = {{More}, Surhud and {van den Bosch}, Frank C. and {Cacciato}, Marcello and {Skibba}, Ramin and {Mo}, H.~J. and {Yang}, Xiaohu},
        title = "{Satellite kinematics - III. Halo masses of central galaxies in SDSS}",
      journal = {\mnras},
         year = 2011,
        month = jan,
       volume = {410},
       number = {1},
        pages = {210-226},
          doi = {10.1111/j.1365-2966.2010.17436.x},
archivePrefix = {arXiv},
       eprint = {1003.3203},
 primaryClass = {astro-ph.CO},
       adsurl = {https://ui.adsabs.harvard.edu/abs/2011MNRAS.410..210M}
}

@ARTICLE{Velander2014,
       author = {{Velander}, Malin and {van Uitert}, Edo and {Hoekstra}, Henk and {Coupon}, Jean and {Erben}, Thomas and {Heymans}, Catherine and {Hildebrandt}, Hendrik and {Kitching}, Thomas D. and {Mellier}, Yannick and {Miller}, Lance and {Van Waerbeke}, Ludovic and {Bonnett}, Christopher and {Fu}, Liping and {Giodini}, Stefania and {Hudson}, Michael J. and {Kuijken}, Konrad and {Rowe}, Barnaby and {Schrabback}, Tim and {Semboloni}, Elisabetta},
        title = "{CFHTLenS: the relation between galaxy dark matter haloes and baryons from weak gravitational lensing}",
      journal = {\mnras},
         year = 2014,
        month = jan,
       volume = {437},
       number = {3},
        pages = {2111-2136},
          doi = {10.1093/mnras/stt2013},
archivePrefix = {arXiv},
       eprint = {1304.4265},
 primaryClass = {astro-ph.CO},
       adsurl = {https://ui.adsabs.harvard.edu/abs/2014MNRAS.437.2111V}
}

@ARTICLE{wojtak13,
       author = {{Wojtak}, Rados{\l}aw and {Mamon}, Gary A.},
        title = "{Physical properties underlying observed kinematics of satellite galaxies}",
      journal = {\mnras},
         year = 2013,
        month = jan,
       volume = {428},
       number = {3},
        pages = {2407-2417},
          doi = {10.1093/mnras/sts203},
archivePrefix = {arXiv},
       eprint = {1207.1647},
 primaryClass = {astro-ph.CO},
       adsurl = {https://ui.adsabs.harvard.edu/abs/2013MNRAS.428.2407W}
}

@ARTICLE{Behroozi19,
       author = {{Behroozi}, Peter and {Wechsler}, Risa H. and {Hearin}, Andrew P. and {Conroy}, Charlie},
        title = "{UNIVERSEMACHINE: The correlation between galaxy growth and dark matter halo assembly from z = 0-10}",
      journal = {\mnras},
         year = 2019,
        month = sep,
       volume = {488},
       number = {3},
        pages = {3143-3194},
          doi = {10.1093/mnras/stz1182},
archivePrefix = {arXiv},
       eprint = {1806.07893},
 primaryClass = {astro-ph.GA},
       adsurl = {https://ui.adsabs.harvard.edu/abs/2019MNRAS.488.3143B}
}

@ARTICLE{Zacharegkas25,
       author = {{Zacharegkas}, G. and {Chang}, C. and {Prat}, J. and {Hartley}, W. and {Mucesh}, S. and {Alarcon}, A. and {Alves}, O. and {Amon}, A. and {Bechtol}, K. and {Becker}, M.~R. and {Bernstein}, G. and {Blazek}, J. and {Campos}, A. and {Carnero Rosell}, A. and {Carrasco Kind}, M. and {Cawthon}, R. and {Chen}, R. and {Choi}, A. and {Cordero}, J. and {Davis}, C. and {Derose}, J. and {Diehl}, H. and {Dodelson}, S. and {Doux}, C. and {Drlica-Wagner}, A. and {Eckert}, K. and {Eifler}, T.~F. and {Elvin-Poole}, J. and {Everett}, S. and {Fang}, X. and {Ferte}, A. and {Gatti}, M. and {Giannini}, G. and {Gruen}, D. and {Gruendl}, R.~A. and {Harrison}, I. and {Huang}, H. and {Huff}, E.~M. and {Jarvis}, M. and {Krause}, E. and {Kuropatkin}, N. and {Leget}, P.~F. and {Maccrann}, N. and {McCullough}, J. and {Myles}, J. and {Alsina}, A.~N. and {Pandey}, S. and {Raveri}, M. and {Rollins}, R.~P. and {Roodman}, A. and {Ross}, A.~J. and {Rykoff}, E.~S. and {Sanchez}, C. and {Secco}, L.~F. and {Sevilla-Noarbe}, I. and {Sheldon}, E. and {Shin}, T. and {Troxel}, M.~A. and {Tutusaus}, I. and {Yanny}, B. and {Yin}, B. and {Zhang}, Y. and {Zuntz}, J. and {Aguena}, M. and {Andrade-Oliveira}, F. and {Bacon}, D. and {Brooks}, D. and {Burke}, D.~L. and {Carretero}, J. and {Castander}, F.~J. and {da Costa}, L.~N. and {da Silva Pereira}, M.~E. and {Davis}, T.~M. and {De Vicente}, J. and {Flaugher}, B. and {Frieman}, J. and {Garcia-Bellido}, J. and {Gaztanaga}, E. and {Gutierrez}, G. and {Hinton}, S.~R. and {Hollowood}, D.~L. and {James}, D.~J. and {Kuehn}, K. and {Lahav}, O. and {Lee}, S. and {Marshall}, J.~L. and {Mena-Fernandez}, J. and {Miquel}, R. and {Mohr}, J.~J. and {Ogando}, R.~L.~C. and {Plazas Malagon}, A.~A. and {Porredon}, A. and {Samuroff}, S. and {Sanchez}, E. and {Smith}, M. and {Soares-Santos}, M. and {Suchyta}, E. and {Swanson}, M.~E.~C. and {Tucker}, D.~L. and {Vikram}, V. and {Weaverdyck}, N. and {Wiseman}, P. and {Yamamoto}, M.},
        title = "{Constraining the Stellar-to-Halo Mass Relation with Galaxy Clustering and Weak Lensing from DES Year 3 Data}",
      journal = {arXiv e-prints},
         year = 2025,
        month = jun,
          eid = {arXiv:2506.22367},
        pages = {arXiv:2506.22367},
          doi = {10.48550/arXiv.2506.22367},
archivePrefix = {arXiv},
       eprint = {2506.22367},
 primaryClass = {astro-ph.GA},
       adsurl = {https://ui.adsabs.harvard.edu/abs/2025arXiv250622367Z}
}

@ARTICLE{hopkins11,
       author = {{Hopkins}, Philip F. and {Quataert}, Eliot and {Murray}, Norman},
        title = "{Self-regulated star formation in galaxies via momentum input from massive stars}",
      journal = {\mnras},
         year = 2011,
        month = oct,
       volume = {417},
       number = {2},
        pages = {950-973},
          doi = {10.1111/j.1365-2966.2011.19306.x},
archivePrefix = {arXiv},
       eprint = {1101.4940},
 primaryClass = {astro-ph.CO},
       adsurl = {https://ui.adsabs.harvard.edu/abs/2011MNRAS.417..950H}
}

@ARTICLE{Ceverino14,
       author = {{Ceverino}, Daniel and {Klypin}, Anatoly and {Klimek}, Elizabeth S. and {Trujillo-Gomez}, Sebastian and {Churchill}, Christopher W. and {Primack}, Joel and {Dekel}, Avishai},
        title = "{Radiative feedback and the low efficiency of galaxy formation in low-mass haloes at high redshift}",
      journal = {\mnras},
         year = 2014,
        month = aug,
       volume = {442},
       number = {2},
        pages = {1545-1559},
          doi = {10.1093/mnras/stu956},
archivePrefix = {arXiv},
       eprint = {1307.0943},
 primaryClass = {astro-ph.CO},
       adsurl = {https://ui.adsabs.harvard.edu/abs/2014MNRAS.442.1545C}
}

@ARTICLE{fabian12,
       author = {{Fabian}, A.~C.},
        title = "{Observational Evidence of Active Galactic Nuclei Feedback}",
      journal = {\araa},
         year = 2012,
        month = sep,
       volume = {50},
        pages = {455-489},
          doi = {10.1146/annurev-astro-081811-125521},
archivePrefix = {arXiv},
       eprint = {1204.4114},
 primaryClass = {astro-ph.CO},
       adsurl = {https://ui.adsabs.harvard.edu/abs/2012ARA&A..50..455F}
}

@ARTICLE{yang12,
       author = {{Yang}, Xiaohu and {Mo}, H.~J. and {van den Bosch}, Frank C. and {Zhang}, Youcai and {Han}, Jiaxin},
        title = "{Evolution of the Galaxy-Dark Matter Connection and the Assembly of Galaxies in Dark Matter Halos}",
      journal = {\apj},
         year = 2012,
        month = jun,
       volume = {752},
       number = {1},
          eid = {41},
        pages = {41},
          doi = {10.1088/0004-637X/752/1/41},
archivePrefix = {arXiv},
       eprint = {1110.1420},
 primaryClass = {astro-ph.CO},
       adsurl = {https://ui.adsabs.harvard.edu/abs/2012ApJ...752...41Y}
}

@ARTICLE{yang03,
       author = {{Yang}, Xiaohu and {Mo}, H.~J. and {van den Bosch}, Frank C.},
        title = "{Constraining galaxy formation and cosmology with the conditional luminosity function of galaxies}",
      journal = {\mnras},
         year = 2003,
        month = mar,
       volume = {339},
       number = {4},
        pages = {1057-1080},
          doi = {10.1046/j.1365-8711.2003.06254.x},
archivePrefix = {arXiv},
       eprint = {astro-ph/0207019},
 primaryClass = {astro-ph},
       adsurl = {https://ui.adsabs.harvard.edu/abs/2003MNRAS.339.1057Y}
}

@ARTICLE{bryan98,
       author = {{Bryan}, Greg L. and {Norman}, Michael L.},
        title = "{Statistical Properties of X-Ray Clusters: Analytic and Numerical Comparisons}",
      journal = {\apj},
         year = 1998,
        month = mar,
       volume = {495},
       number = {1},
        pages = {80-99},
          doi = {10.1086/305262},
archivePrefix = {arXiv},
       eprint = {astro-ph/9710107},
 primaryClass = {astro-ph},
       adsurl = {https://ui.adsabs.harvard.edu/abs/1998ApJ...495...80B}
}

@ARTICLE{Chabrier03,
       author = {{Chabrier}, Gilles},
        title = "{Galactic Stellar and Substellar Initial Mass Function}",
      journal = {\pasp},
         year = 2003,
        month = jul,
       volume = {115},
       number = {809},
        pages = {763-795},
          doi = {10.1086/376392},
archivePrefix = {arXiv},
       eprint = {astro-ph/0304382},
 primaryClass = {astro-ph},
       adsurl = {https://ui.adsabs.harvard.edu/abs/2003PASP..115..763C}
}

@ARTICLE{Calzetti00,
       author = {{Calzetti}, Daniela and {Armus}, Lee and {Bohlin}, Ralph C. and {Kinney}, Anne L. and {Koornneef}, Jan and {Storchi-Bergmann}, Thaisa},
        title = "{The Dust Content and Opacity of Actively Star-forming Galaxies}",
      journal = {\apj},
         year = 2000,
        month = apr,
       volume = {533},
       number = {2},
        pages = {682-695},
          doi = {10.1086/308692},
archivePrefix = {arXiv},
       eprint = {astro-ph/9911459},
 primaryClass = {astro-ph},
       adsurl = {https://ui.adsabs.harvard.edu/abs/2000ApJ...533..682C}
}

@ARTICLE{Yesuf2026,
       author = {{Yesuf}, Hassen M. and {Joshi}, Ravi and {Zou}, Yuxuan and {Yuan}, Feng and {Ho}, Luis C. and {Lin}, Lin and {Hao}, Lei and {Shen}, Shiyin and {Bottrell}, Connor and {Guo}, Fulai and {Silverman}, John D.},
        title = "{Widespread Inflows Reveal Baryonic Cycling in Star-forming and Quiescent Galaxies}",
      journal = {arXiv e-prints},
         year = 2026,
        month = sep,
          eid = {arXiv:2609.04327},
        pages = {arXiv:2609.04327},
          doi = {10.48550/arXiv.2609.04327},
archivePrefix = {arXiv},
       eprint = {2609.04327},
 primaryClass = {astro-ph.GA},
       adsurl = {https://ui.adsabs.harvard.edu/abs/2026arXiv260904327Y}
}

@ARTICLE{Bruzual2003,
       author = {{Bruzual}, G. and {Charlot}, S.},
        title = "{Stellar population synthesis at the resolution of 2003}",
      journal = {\mnras},
         year = 2003,
        month = oct,
       volume = {344},
       number = {4},
        pages = {1000-1028},
          doi = {10.1046/j.1365-8711.2003.06897.x},
archivePrefix = {arXiv},
       eprint = {astro-ph/0309134},
 primaryClass = {astro-ph},
       adsurl = {https://ui.adsabs.harvard.edu/abs/2003MNRAS.344.1000B}
}

@Misc{methods,
  note = {Materials and methods are available as supplementary material},
}

\end{document}